\documentclass[11pt]{article}
\usepackage{epstopdf}
\usepackage{color}
\usepackage{subfigure}
\usepackage{amsmath}
\usepackage{amssymb}
\usepackage{graphicx,color}
\usepackage{cite}
\usepackage{enumerate}
\usepackage{amsthm}
\usepackage{amsfonts,mathrsfs}
\usepackage{geometry}
\usepackage{ifthen}
\usepackage{graphicx}
\usepackage{float}
\usepackage{bm}
\usepackage{tabularx}
\usepackage[utf8]{inputenc}
\usepackage{amsfonts}
\usepackage[colorlinks,bookmarksopen,bookmarksnumbered,citecolor=blue, linkcolor=blue, urlcolor=blue]{hyperref}
\usepackage{booktabs}
\usepackage{array}

\begin{document}

\title{\bf Quantifying imaginarity of quantum operations}

\vskip0.1in
\author{\small Chuanfa Wu$^1$, Zhaoqi Wu$^1$\thanks{Corresponding author. E-mail: wuzhaoqi\_conquer@163.com}\\
{\small\it  1. Department of Mathematics, Nanchang University, Nanchang 330031, P R China}\\
}

\date{}
\maketitle

\noindent {\bf Abstract} {\small }\\
\indent Complex numbers are theoretically proved and experimentally
confirmed as necessary in quantum mechanics and quantum information,
and a resource theory of imaginarity of quantum states has been
established. In this work, we establish a framework to quantify the
imaginarity of quantum operations from the perspective of the
ability to create or detect imaginarity, following the idea by
Theurer {\it et al.} [Phys. Rev. Lett. \textbf{122}, 190405 (2019)]
used in coherence theory. We introduce two types of imaginarity
measures of quantum operations based on the norm and the weight,
investigate their properties and relations, derive the analytical
formulas of the measure under the trace norm for qubit unitary
operations, and present some applications in the tasks of channel
discrimination and the entanglement-assisted exclusion. The results
provide new insights into imaginarity of operations and deepen our
understanding of dynamical imaginarity.


\noindent {\bf 1 Introduction} \\ \hspace*{\fill}\\
\indent Quantum mechanics is a basic theory describing the microscopic world. It reveals the behavioral laws of microscopic particles such as atoms\cite{CY,SHP,LERF} and molecules\cite{WRG,EPWP}, and explains phenomena that classical physics cannot describe, for instance, the photoelectric effect\cite{KS,NMKS} and atomic spectra\cite{BHAS,JBR}. The most distinctive feature of quantum mechanics is that imaginary numbers are no longer merely mathematical tools but experimentally necessary. In fact, since the birth of quantum mechanics, there has been ongoing debate about whether complex numbers are truly essential in quantum mechanics. Owing to the fact that physical experiments are described using probabilities, i.e., real numbers, some early researchers argued that complex numbers were not theoretically necessary in quantum mechanics and were introduced merely to simplify calculations\cite{BGVN,SEC,MMMM,DFJ,AABV}. However, Renou et al.\cite{RMOT} designed an experiment in an entanglement-swapping scenario and found that the predictions of complex-valued quantum mechanics matched the experimental results, while those of real-valued quantum mechanics deviated. This demonstrates that complex numbers are indispensable in quantum mechanics. Recently, Wu et al. \cite{WDJY} performed an experiment under strict locality conditions, demonstrating that complex-valued quantum mechanics is more complete and accurate in describing experimental phenomena compared to real-valued ones.

To systematically study and quantify the value of complex numbers in
quantum systems, as well as the potential advantages they may offer,
such as nonlocality\cite{WZFS} and discrimination
tasks\cite{ZFPN,BABJ,BJBT}, in 2018, Hickey and Gour\cite{HAGG}
proposed to treat the imaginarity of quantum states as a quantum
resource and provided a framework for the imaginarity resource
theory. In general resource theories\cite{BFGG,CBFT,CEGG}, there are
two essential elements: free states and free operations. The free
states in imaginarity resource theory are real states, i.e., the
quantum states satisfying $\langle i|\rho |j\rangle \in \mathbb{R}$
for any $i$ and $j$, in which $\{|i\rangle\}$ is a given orthonormal
basis in a Hilbert space, and free operations are those that can map
real states to real states. The most trivial free operations are
those with real Kraus operators (RKO), i.e., $\langle m|K_{j}| n
\rangle \in \mathbb {R} \notag $ for arbitrary $j,m$, and $n$, where
$K_j$ are the Kraus operators of a quantum operation $\Lambda$. Here
$\mathbb {R}$ denotes the set of all real numbers.

At present, many studies focus on quantifying the imaginarity of
quantum states with different physical meanings, such as the $l_1$
norm of imaginarity\cite{CQGT}, the robustness of
imaginarity\cite{HAGG,WKKT}, the fidelity of imaginarity\cite{WKKT},
the geometric-like imaginarity\cite{GMLB}, and the unified
$(\alpha,\beta)$-relative entropy of imaginarity\cite{WCWZ}. On the
other hand, the manipulation of imaginarity resource, i.e., the
problem of whether any two given imaginary states can be converted
into each other via free operations, has been extensively
studied\cite{HAGG,WKKT,GMLB,DSBZ}. Additionally, a series of
imaginarity witnesses\cite{FCWR,ZLLN} and Bargmann invariants for
imaginarity\cite{LMTY,ZLXB} have been proposed, which deepen the
understanding of the characteristics and essence of imaginarity. The intrinsic connections and distinctions between
imaginarity and entanglement\cite{SYRR}/imaginarity and
coherence\cite{XJ25,ZLLN24,LHHM} have been clarified. Similar to
other quantum resources\cite{SUKC,BTRC}, issues like broadcasting of
imaginarity\cite{ZZLN} and freezing imaginarity\cite{HSZB} have also
been investigated.

In coherence resource theory, Baumgratz et al.\cite{BTCM} first
established a framework for coherence of quantum states.
Subsequently, an equivalent yet simplified framework\cite{YXZD} has
been proposed, which significantly advanced this field. So far,
remarkable progress has been made in quantifying
coherence\cite{TKCV,NCBT,WZZL,FJWZ,XJ16}, characterizing state
transformations\cite{DSBZ15,BKSU,WKTT}, and exploring operational
advantages in quantum information processing\cite{MDPT,MZCJ,DSMA}.
Coherence resource theories have been established not only for
quantum states but also for quantum operations. Xu \cite{XJ19} used
the Choi-Jamio\l kowski isomorphism to establish a one-to-one
correspondence between quantum operations and quantum states,
thereby transforming the study of coherence of quantum operations
into the study of coherence of quantum states. The coherence of
quantum operations defined above depends on coherent states.
Alternatively, the coherence of quantum operations can also be
quantified by their ability to generate or detect
coherence\cite{TTED}, which do not rely on coherent states. In
imaginarity resource theory, Chen and Lei \cite{CXLQ} investigated
the quantum imaginarity of operations by Choi-Jamio\l kowski
isomorphism. In this paper, we propose a framework
of imaginarity of quantum operations and some quantifiers from the
perspective of the ability to create or detect imaginarity, which is
different from the regime based on Choi-Jamio\l kowski isomorphism.

The remainder of this paper is organized as follows.
In Section 2, we introduce some necessary notations and review some
related concepts. In Section 3, we establish the framework of
imaginarity resource theory of quantum operations. In Section 4, we
present two classes of imaginarity quantifiers of quantum operations
based on the norm and the weight, explore their properties and build
the relations between them. In Section 5, we give an explicit
example to illustrate our results. In Section 6, we present the
applications of imaginarity as a quantum resource in the channel
discrimination task and the entanglement-assisted exclusion task.
The conclusions are provided in Section 7.\\

\noindent {\bf   2 Preliminaries}\\\hspace*{\fill}

Let $\mathcal{H}$ be a $d$-dimensional Hilbert space, and $\{|i\rangle \}_{i=0}^{d-1}$ an orthonormal basis for $\mathcal {H}$. Specifically, we denote by $\mathcal{H_A}$ and $\mathcal{H_B}$ two Hilbert spaces with dimensions $|A|$ and $|B|$, and $\{|a_i\rangle \}_{i=0}^{|A|-1}$ and $\{|b_k\rangle \}_{k=0}^{|B|-1}$ the orthonormal bases for $\mathcal {H_A}$ and $\mathcal {H_B}$, respectively. Denote by $\mathcal D( \mathcal {H_A}) $ and $\mathcal D( \mathcal {H_B})$ the sets of density operators (quantum states) acting on $  \mathcal {H_A} $ and $  \mathcal {H_B}$, respectively, and $\mathcal{O_{AB}}$ the set of all quantum operations from $\mathcal L( \mathcal {H_A})$ to $\mathcal L( \mathcal {H_B})$, where $\mathcal{L}(X)$ denotes the set of all linear operators on $X$. Throughout this paper, we denote by $\mathbb{I}$ the identity operation, $\mathbb{R}$ the set of all real numbers, and $\mathbb{Z^+}$ the set of all positive integers. In addition, if we concatenate operations, we always assume that the output dimension of the first operation equals to the input dimension of the second operation and we will omit the concatenation operator $\circ$ if not necessary. We recall some basic concepts.\\
  \indent A quantum operation $\Phi$ is a completely positive map with the Kraus representation $\Phi(\rho)=\sum\limits_{n}K_n \rho K_{n}^{\dagger}$, in which $\{K_n\}$ is a set of Kraus operators from $\mathcal{H_A}$ into $\mathcal{H_B}$ satisfying $\sum\limits_{n} K_{n}^{\dagger}K_{n}\le \mathbf{I}$, where $\mathbf{I}$ denotes the identity operator. A quantum operation $\Phi$ is called a quantum channel if it is trace preserving, i.e., $\sum\limits_{n} K_{n}^{\dagger}K_{n}=\mathbf{I}$.

 \indent  {\bf Definition 1}\cite{CQGT} For any $\rho\in\mathcal{D}(\mathcal{H})$, the deimaginarity map $\bigtriangleup $ is defined by
 \begin{equation}\label{eq1}
    \bigtriangleup(\rho)=\dfrac{1}{2}\left(\rho+\rho^{\mathrm{T}}\right)=\sum_{ij}\mathrm{Re}\left(\rho_{ij}\right)|i\rangle \langle j|
 \end{equation} where $\rho=\sum\limits_{ij}\rho_{ij}|i\rangle \langle j|$, $\mathrm{Re}\left(\rho_{ij}\right)$ represents the real part of $\rho_{ij}$, and $\rho^T$ denotes the transposition of $\rho$. Its output is always a real state w.r.t. $\{|i\rangle \}_{i=0}^{d-1}$.

  It bears noting that $\bigtriangleup $ is a positive map, but not a completely positive map, that is, $\bigtriangleup$ is not a quantum operation\cite{CQGT}. To see this, consider $\rho=(|00\rangle \langle00|+|00\rangle \langle 11|+ |11\rangle \langle00|+|11\rangle \langle 11|)/4$. Then we have
 \begin{align}
 (\bigtriangleup \otimes \mathbb{I}_2)(\rho)=
 \begin{pmatrix}
    1 & 0 & 0 & \frac{1}{2} \\
    0 & 0 & \frac{1}{2} & 0 \\
    0 & \frac{1}{2} & 0 & 0 \\
    \frac{1}{2} & 0 & 0 & 1 \\
 \end{pmatrix}, \notag
 \end{align}
which is not semidefinite. Here $\bigtriangleup$ is the
deimaginarity map on $\mathcal{D}(\mathbb{C}^2)$, $\mathbb{I}_2$ denotes the
identity map on $\mathcal{D}(\mathbb{C}^2)$ and
$\rho\in\mathcal{D}(\mathbb{C}^2\otimes \mathbb{C}^2)$. In the
following, we denote by $\bigtriangleup^{A}$, $\bigtriangleup^B$, and
$\bigtriangleup^{AB}$ the deimaginarity maps on $\mathcal{H_A}$,
$\mathcal{H_B}$, and $\mathcal{H_A} \otimes \mathcal{H_B}$,
respectively.

 In resource theory, norms are often used to induce resource measures, such as coherence measures based on $l_1$-norm\cite{BTCM} and trace norm\cite{SLXZ,RSPP}. To construct imaginarity measures of quantum operations using norms, we first recall some concepts.

The norm $\lVert \Psi \rVert$ of a superoperator $\Psi : \mathcal L( \mathcal {H_A}) \rightarrow \mathcal L( \mathcal {H_B}) $ is defined
by \cite{PVUV}
 \begin{align} \label{eq2}
    \lVert \Psi \rVert=\max \left\{ \lVert\Psi(\rho)\rVert: \rho \in \mathcal{D(H_A)}, \lVert \rho \rVert \le 1 \right  \}.
 \end{align}
The above norm $\lVert \cdot \rVert$ is called submultiplicative if
 \begin{align}\label{eq3}
    \lVert \Psi_1 \circ \Psi_2 \rVert \le   \lVert \Psi_1\rVert  \lVert \Psi_2 \rVert,
 \end{align} and submultiplicative under tensor product if
 \begin{align} \label{eq4}
    \lVert \Psi_1 \otimes \Psi_2 \rVert \le     \lVert \Psi_1\rVert  \lVert \Psi_2 \rVert
 \end{align} for any superoperators $\Psi_1$ and $\Psi_2$ \cite{PVUV}.

For any operator $A$, the Schatten $p$-norm of $A$ is defined by \cite{WATJ}
 \begin{align}\label{eq5}
 \lVert A \rVert_p = \left( \mathrm{tr}|A|^{p}\right)^\frac{1}{p},
 \end{align} where $|A|=\sqrt{A^{\dagger}A}$ and $p\ge 1$.\\

\noindent {\bf 3 A framework for quantifying imaginarity of quantum operations}\\\hspace*{\fill}\\
\indent In order to establish the framework of imaginarity resource theory of quantum operations, we need to specify the definitions of free operations and free superoperations.\\
\indent Let us begin with free POVMs. A positive operator-valued
measure (POVM) $E=\{E_n\}$ is a set of operators satisfying $E_n \ge
0$ and $\sum\limits_{n}{E_n}=\mathbf{I}$. We present the definition
of a free POVM as follows.

\indent {\bf Definition 2} A POVM $E=\{E_n\}$ on $\mathcal{H_A}$ is free if
\begin{equation}\label{eq6}
    \mathrm{tr}\left( E_n \bigtriangleup (\rho)\right) = \mathrm{tr} \left( E_n \rho\right)
\end{equation} for any $\rho \in \mathcal{D(H_A)}$ and $n$.

\indent It follows from Definition 2 that a POVM cannot detect imaginarity if the probabilities of the measure outcome are independent of them. Then a significant question is how to characterize free POVMs. This leads to the following theorem.

 \indent {\bf Theorem 1} A POVM is free if and only
if
\begin{align}\label{eq7}
    E_n=\sum_{i,j}\mathrm{Re}\left(E_{ij}^{n}\right)|a_i\rangle \langle a_j|,
\end{align} namely $E_{n}^{\mathrm{T}}=E_{n}$ for any $n$, where $E_{ij}^{n}=\langle a_j|E_n|a_i\rangle$ and $E_{n}^{\mathrm{T}}$ denotes the transposition of $E_{n}$.\\
\indent \textit {Proof}. Let $E_n=\sum_{i,j}\limits E_{ij}^{n}|a_i\rangle \langle a_j|$ and $\rho=\sum\limits_{s,t}\rho_{st}|a_s\rangle \langle a_t|$. Then we obtain
\begin{align}
    \mathrm{tr}(E_n\rho)=\sum_{i,j}E_{ij}^n\rho_{ji} \notag
\end{align} and
\begin{align}
    \mathrm{tr}\left( E_n\bigtriangleup(\rho)\right) =\sum_{i,j}E_{ij}^n\mathrm{Re}\left( \rho_{ji}\right). \notag
\end{align} Therefore,
\begin{align}
    \mathrm{tr}(E_n\rho)=\mathrm{tr}\left( E_n\bigtriangleup(\rho)\right)\Leftrightarrow \sum_{i,j}E_{ij}^n\left[\rho_{ji}-\mathrm{Re}\left( \rho_{ji}\right)\right]=0. \notag
\end{align} Since $E_n$ and $\rho$ are hermitian for each $n$, we get
\begin{align}
    E_{ij}^n\left[\rho_{ji}-\mathrm{Re}\left( \rho_{ji}\right)\right]+E_{ji}^n\left[\rho_{ij}-\mathrm{Re}\left(\rho_{ij}\right)\right]=2\mathrm{Re}\left( E_{ij}^n\left[\rho_{ji}-\mathrm{Re}\left( \rho_{ji}\right)\right]\right),  \notag
\end{align}
which yields that
\begin{align}
    \sum_{i,j}E_{ij}^n\left[\rho_{ji}-\mathrm{Re}\left( \rho_{ji}\right)\right]=\sum_{i,j}\mathrm{Re}\left( E_{ij}^n\left[\rho_{ji}-\mathrm{Re}\left( \rho_{ji}\right)\right]\right)=2\sum_{i<j}\mathrm{Re}\left(
E_{ij}^n\left[\rho_{ji}-\mathrm{Re}\left(\rho_{ji}\right)\right]\right).\notag
\end{align}
Noting that $\rho_{ji}-\mathrm{Re}\left( \rho_{ji}\right)$ is a pure
imaginary number, it is easy to verify that for each $n$,
$\sum\limits_{i<j}\mathrm{Re}\left(
E_{ij}^n\left[\rho_{ji}-\mathrm{Re}\left(
\rho_{ji}\right)\right]\right)=0$ for any $\rho\in\mathcal{D(H_A)}$
iff $E_{ij}^n\in\mathbb{R}$ for $i<j$. Since $E_n$ are hermitian for
each $n$, we have $E_{ii}^n\in\mathbb{R}$. Hence,
$\mathrm{tr}(E_n\rho)=\mathrm{tr}\left(
E_n\bigtriangleup(\rho)\right)\Leftrightarrow E_{ij}^n\in\mathbb{R}$
for all $i,j$ and $n$, from which the conclusion follows.
$\hfill\qedsymbol$
\\\hspace*{\fill}

\indent Inspired by \cite{TTED}, we now define three types of free operations by considering the ability of detecting or creating imaginarity.

\indent {\bf Definition 3} Let $\Phi\in \mathcal{O_{AB}}$. Then a quantum operation $\Phi$ is called detection real if
\begin{align}\label{eq8}
    \bigtriangleup \Phi =\bigtriangleup \Phi    \bigtriangleup,
\end{align} $\Phi$ is called creation real if
\begin{align}\label{eq9}
     \Phi \bigtriangleup=\bigtriangleup \Phi   \bigtriangleup,
\end{align} and $\Phi$ is called detection creation real if
\begin{align}\label{eq10}
    \bigtriangleup \Phi = \Phi \bigtriangleup.
\end{align} We denote the set of detection real operations, creation real operations and detection creation real operations by DR, CR and DCR, respectively.

\indent In general, a quantum operation $\Phi \in \mathcal{O_{AB}}$ can be represented as
\begin{align}\label{eq11}
    \Phi\left(|a_i\rangle \langle a_j|\right)=\sum_{k,l}\Phi_{k,l}^{i,j}|b_k\rangle \langle b_l|.
\end{align} Therefore, $\Phi$ is completely determined by the coefficients $\Phi_{k,l}^{i,j}$.\\
\indent {\bf Theorem 2} Let $\Phi\in \mathcal{O_{AB}}$. Then following statements are equivalent:\\
\indent (1) $\Phi$ is a DR operation;\\
\indent (2) $\Phi$ is a CR operation;\\
\indent (3) $\Phi$ is a DCR operation;\\
\indent (4) The coefficients $\Phi_{k,l}^{i,j}$ of $\Phi$ in Eq. (\ref{eq11}) are real numbers for any $i,j,k$ and $l$.\\
\indent \textit {Proof}. We prove that (4) is a sufficient and necessary condition for (1), (2) and (3).

Note that
\begin{align}
    \left( \bigtriangleup \Phi\right) (\rho)&=\left( \bigtriangleup \Phi\right) \left( \sum_{i,j}\rho_{ij}|a_i\rangle \langle a_j|\right)  \notag \\
    &=\bigtriangleup \left( \sum_{i,j} \rho_{ij} \Phi(|a_i\rangle \langle a_j|)\right)  \notag \\
    &=\bigtriangleup \left( \sum_{i,j,k,l} \rho_{ij} \Phi_{k,l}^{i,j} |b_k\rangle \langle b_l|\right)  \notag \\
    &=\sum_{i,j,k,l} \mathrm{Re}\left( \rho_{ij} \Phi_{k,l}^{i,j}\right) |b_k\rangle \langle b_l|, \notag
\end{align}
\begin{align}
    \left( \Phi \bigtriangleup\right)  (\rho)&=\Phi \left( \sum_{i,j}\mathrm{Re}(\rho_{ij})|a_i\rangle \langle b_j|\right) \notag \\
    &=\sum_{i,j}\mathrm{Re}(\rho_{ij}) \Phi(|a_i\rangle \langle a_j|) \notag \\
    &=\sum_{i,j,k,l} \mathrm{Re}(\rho_{ij})\Phi_{k,l}^{i,j}|b_k\rangle \langle b_l| \notag
\end{align} and
\begin{align}
    \left( \bigtriangleup \Phi \bigtriangleup\right) (\rho)&=\left( \bigtriangleup \Phi\right)  \left( \sum_{i,j} \mathrm{Re}(\rho_{ij})|a_i\rangle \langle a_j|\right) \notag \\
    &=\bigtriangleup \left( \sum_{i,j,k,l} \mathrm{Re} (\rho_{ij}) \Phi_{k,l}^{i,j} |b_k\rangle \langle b_l|\right)  \notag \\
    &=\sum_{i,j,k,l} \mathrm{Re}\left( \mathrm{Re} (\rho_{ij}) \Phi_{k,l}^{i,j}\right) |b_k\rangle \langle b_l| \notag \\
    &=\sum_{i,j,k,l} \mathrm{Re} \left( \rho_{ij}\right)  \mathrm{Re}\left( \Phi_{k,l}^{i,j}\right) |b_k\rangle \langle b_l|. \notag
\end{align}

Firstly, $\Phi$ is a DR operation iff $\left( \bigtriangleup \Phi\right) (\rho)= \left( \bigtriangleup \Phi \bigtriangleup\right) (\rho)$ for any $\rho$, iff $ \sum \limits_{i,j,k,l} \mathrm{Re}\left( \rho_{ij} \Phi_{k,l}^{i,j}\right)\\ |b_k\rangle \langle b_l|=\sum\limits_{i,j,k,l} \mathrm{Re} \left( \rho_{ij}\right)  \mathrm{Re}\left( \Phi_{k,l}^{i,j}\right) |b_k\rangle \langle b_l|$ for any $\rho_{ij}$, iff $ \Phi_{k,l}^{i,j} \in \mathbb{R}$ for any $i,j,k$ and $l$. Therefore, item (1) and item (4) are equivalent.

Secondly, $\Phi$ is a CR operation iff $\left( \Phi \bigtriangleup\right)  (\rho)=\left( \bigtriangleup \Phi \bigtriangleup\right) (\rho)$ for any $\rho$, iff $ \sum\limits_{i,j,k,l} \mathrm{Re}(\rho_{ij})\Phi_{k,l}^{i,j}\\|b_k\rangle \langle b_l|=\sum\limits_{i,j,k,l} \mathrm{Re} \left( \rho_{ij}\right)  \mathrm{Re}\left( \Phi_{k,l}^{i,j}\right) |b_k\rangle \langle b_l|$ for any $\rho_{ij}$, iff $\Phi_{k,l}^{i,j} \in \mathbb{R}$ for any $i,j,k$ and $l$. It means that item (2) and item (4) are equivalent.

Finally, $\Phi$ is a DCR operation iff $\left( \bigtriangleup \Phi\right)  (\rho)=\left( \Phi \bigtriangleup\right)  (\rho)$ for any $\rho$, iff $ \sum\limits_{i,j,k,l} \mathrm{Re}\left( \rho_{ij} \Phi_{k,l}^{i,j}\right) \\ |b_k\rangle \langle b_l|=\sum\limits_{i,j,k,l} \mathrm{Re}(\rho_{ij})\Phi_{k,l}^{i,j}|b_k\rangle \langle b_l| $ for any $\rho_{ij}$, iff $\Phi_{k,l}^{i,j} \in \mathbb{R}$ for any $i,j,k$ and $l$. This implies that item (3) and item (4) are equivalent. Therefore, the statements are all equivalent. $\hfill\qedsymbol$ \\\hspace*{\fill}\\
\indent Note that the largest set of free operations is RNG
(resource non-generating) operations. It follows from Corollary 1 in
Ref.\cite{HAGG} that  $\Phi$ is physically consistent iff
$\Phi(\rho)^{\mathrm{T}}=\Phi(\rho^{\mathrm{T}})$ for any $\rho \in \mathcal{D(H_A)}$. By
Theorem 3 in \cite{CQGT}, if $\Phi$ is physically consistent, we
have $(\bigtriangleup \Phi)(\rho)= (\Phi\bigtriangleup)(\rho)$ for any
$\rho$, i.e., $\Phi$ is a DCR operation. From its proof, it can be
easily verified that the converse also holds. This indicates that
the set of DCR operations coincides with the one of physically
consistent operations.

For simplicity, we denote by $\mathcal{FO_{AB}}$ the set of
all free operations from $\mathcal L( \mathcal {H_A})$ to $\mathcal
L( \mathcal {H_B})$, i.e., the set of DR/CR/DCR operations. The
following theorem describes the relation between the free operation
$\Phi$ and its corresponding Kraus operators $\{K_n\}$.

\indent {\bf Theorem 3} Suppose $\Phi \in \mathcal{O_{AB}}$ determined by the coefficients $\Phi_{k,l}^{i,j}$ in Eq. (\ref{eq11}) with its Kraus operators $\{K_n\}$. Then $\Phi \in \mathcal{FO_{AB}}$ if and only if
\begin{align} \label{eq12}
    \sum_{n}\left( K_n \right)_{k,i}  \left( K^*_{n} \right)_{l,j} \in \mathbb{R}
\end{align} for arbitrary $i$, $j \in \{0,1,\cdots , |A|-1\}$ and $k$, $l\in \{0,1, \cdots, |B|-1\}$, where $\left( K_n \right)_{k,i}=\langle i|K_n|k\rangle$ and $K^*_{n} $ is the conjugate of $K_{n}$.

\indent \textit {Proof}. It follows from \cite{TTED} that $\Phi$ is a quantum operation represented by Eq. (\ref{eq11}) iff $\Phi_{k,l}^{i,j}=\sum\limits_{n}\left( K_n \right)_{k,i}  \left( K^*_{n} \right)_{l,j}$. Using Theorem 2, we obtain that $\Phi \in \mathcal{FO_{AB}}$ iff $ \Phi_{k,l}^{i,j} \in \mathbb{R}$ for any $i,j,k$ and $l$ iff $ \sum\limits_{n}\left( K_n \right)_{k,i}  \left( K^*_{n} \right)_{l,j} \in \mathbb{R}$ for for any $i,j,k$ and $l$. $\hfill\qedsymbol$ \\\hspace*{\fill}\\
\indent  From Theorem 3, it is easy to see that a RKO operation is a
DR/CR/DCR operation. However, the converse is not true in general.
In fact, consider the quantum operation $\tilde{\Theta}$ with Kraus
operators
\begin{equation*}
    K_1=\begin{pmatrix} \frac{1}{4}+\frac{1}{2} \mathrm{i} & 0 \\ 0 & \frac{1}{4}-\frac{1}{2} \mathrm{i} \end{pmatrix}
    \text{ and }
    K_2=\begin{pmatrix} -\frac{\sqrt{2}}{4}+\frac{\sqrt{2}}{4} \mathrm{i} & 0 \\ 0 & -\frac{\sqrt{2}}{4}-\frac{\sqrt{2}}{4} \mathrm{i} \end{pmatrix}.
\end{equation*} It is obvious that $\tilde{\Theta}$ does not belong to RKO operations. But direct calculations show that $K_1$ and $K_2$ satisfy Eq. (\ref{eq12}), i.e., $\tilde{\Theta}$ is a DR/CR/DCR operation.

\begin{figure}[H]\centering
    {\begin{minipage}[figure1]{0.6\linewidth}
            \includegraphics[width=1.0\textwidth]{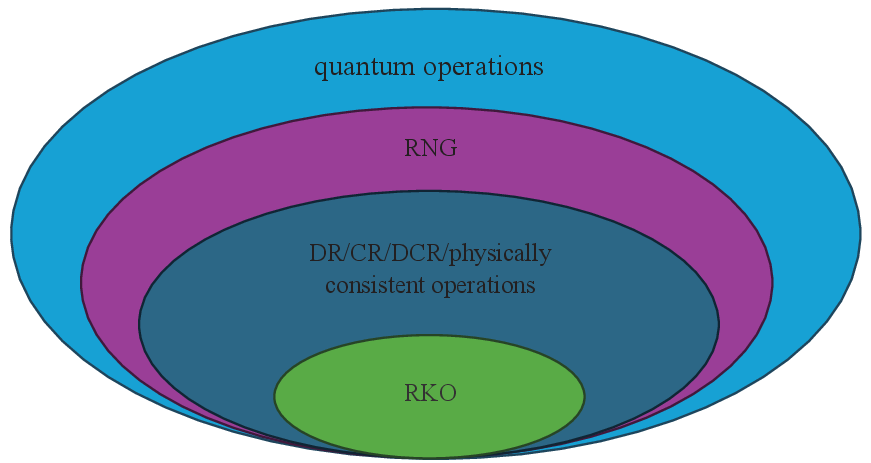}
    \end{minipage}}
    \caption{{A Venn diagram illustrating the relations among different classes of free operations in imaginarity resource theory \label{fig:Fig1}}}
\end{figure}

Based on the above arguments, we depict the relations among
different classes of free operations in imaginarity theory in Figure \ref{fig:Fig1}.

Next, we define free superoperations. The minimum requirement for
free superoperations is that they can map free operations to free
operations.

{\bf Definition 4} For free operation $\Phi$, elementary free superoperations are given by
\begin{align}\label{eq13}
    \mathcal{E}_{1,\Phi}[\Theta]=\Phi \circ \Theta, \qquad \mathcal{E}_{2,\Phi}[\Theta]=\Theta \circ \Phi.
\end{align}  Since the proof of Lemma 11 in \cite{TTED} is independent of the definition of $\bigtriangleup$, we know that $    \mathcal{E}_{1,\Phi}[\Theta]$ and $ \mathcal{E}_{2,\Phi}[\Theta]$ are free operations under the three different settings mentioned above if the input operation $\Theta$ is a free operation. This indicates that the elementary free superoperations we defined are suitable.

A free superoperation $\mathcal{F}$ is the composition of a sequence
of elementary free superoperations, i.e.,
\begin{align}\label{eq14}
    \mathcal{F}= \mathcal{E}_{i_n,\Phi_n}\cdots \mathcal{E}_{i_2,\Phi_2}\mathcal{E}_{i_1,\Phi_1},
\end{align} where $i_j\in \{1,2\}$ and $\Phi_j$ for $j\in \{1,2,\cdots,n\}$, and $n\in\mathbb{Z^+}$. We denote the set of all free superoperations from $\mathcal{FO_{AB}}$ to $\mathcal{FO_{A^{'}B^{'}}}$ by $\mathcal{FSO_{ABA^{'}B^{'}}}$.\\

\noindent {\bf 4 Imaginarity measures of quantum operations}\\\hspace*{\fill}\\
\indent In this section, we define the imaginarity of quantum operations induced by the norm and the weight, and discuss the properties of them and the relations among them.

Analogous to coherence resource theory\cite{HAGG,BTCM,XJ19,TTED},
the  imaginarity measure of quantum operations should satisfy
faithfulness, monotonicity under free superoperation, and convexity.
These requirements lead to the following definition.

{\bf Definition 5} An imaginarity measure of quantum operations is a functional $M$ from $\mathcal{O_{AB}}$ to $[0,+\infty)$  satisfying the following properties:\\
\indent (M1) Faithfulness. $M(\Theta)=0$ if and only if $\Theta \in \mathcal{FO_{AB}}$.\\
\indent (M2) Monotonicity. $M\left( \mathcal{F}[\Theta]\right) \le M(\Theta)$ whenever $\mathcal{F} \in \mathcal{FSO_{ABA^{'}B^{'}}}
$.\\
\indent (M3) Convexity. $M\left( \sum \limits_{j}p_{j} \Theta_{j}\right)  \le \sum \limits_{j}p_{j}M(\Theta_{j})$ for any set of quantum operation $\{\Theta_{j}\}$ and any  $p_{j} \ge 0$ with $\sum \limits_{j}p_{j}=1 $.

It follows from Definition 4 that (M2) is equivalent to (M2a) $+$ (M2b):\\
\indent (M2a) $M(\mathcal{E}_{1,\Phi}[\Theta])\le M(\Theta)$ wherever $\Theta \in \mathcal{O_{AB}}$ and $\Phi \in \mathcal{FO_{BC}}$.\\
\indent (M2b) $M(\mathcal{E}_{2,\Phi}[\Theta])\le M(\Theta)$
wherever $\Theta \in \mathcal{O_{AB}}$ and $\Phi \in
\mathcal{FO_{DA}}$.

\indent Before constructing imaginarity measure of quantum operations, we provide some elegant properties of $\bigtriangleup$, which are crucial for the subsequent proof of the theorems.\\
\indent {\bf Theorem 4} The deimaginarity map $ \bigtriangleup$ has the following properties:\\
\indent (1) $\bigtriangleup = \bigtriangleup^2 $ where $ \bigtriangleup$ is defined by Eq. (\ref{eq1}).  \\
\indent (2) $\bigtriangleup^{AB}(\rho \otimes \sigma)=\bigtriangleup^A(\rho) \otimes \bigtriangleup^B(\sigma)$ where $\rho \in \mathcal D( \mathcal {H_A})$ and $\sigma \in \mathcal D( \mathcal {H_B})$ if and only if at least one of $\rho$ or $\sigma $ is a real state.  \\
\indent (3) $(\bigtriangleup^{A} \otimes \mathbb{I}^B)(\rho \otimes \sigma)=\bigtriangleup^{AB}(\rho \otimes \sigma)$ where $\rho \in \mathcal D( \mathcal {H_A})$ and $\sigma \in \mathcal D( \mathcal {H_B})$ if and only if $\sigma$ is a real state.\\
\indent \textit {Proof}. (1) Item (1) is obvious from Eq. (\ref{eq1}).\\
\indent (2) Let $\rho=\sum\limits_{ij} \rho_{ij}|a_i\rangle \langle a_j|$ and $\sigma=\sum\limits_{kl} \sigma_{kl}|b_k\rangle \langle b_l|$. Then we have
\begin{align}
    \bigtriangleup^{AB}(\rho \otimes \sigma)&=\bigtriangleup^{AB}\left( \sum\limits_{ijkl} \rho_{ij}\sigma_{kl}|a_ib_k\rangle \langle a_jb_l|\right) \notag \\
    &=\sum\limits_{ijkl} \mathrm{Re}(\rho_{ij}\sigma_{kl})|a_ib_k\rangle \langle a_jb_l| \notag
\end{align} and
\begin{align}
    &\bigtriangleup^A(\rho) \otimes \bigtriangleup^B(\sigma)\notag \\
    &=\bigtriangleup^A\left(\sum\limits_{ij} \rho_{ij}|a_i\rangle \langle a_j| \right)  \otimes \bigtriangleup^B\left( \sum\limits_{kl} \sigma_{kl}|b_k\rangle \langle b_l|\right) \notag \\
    &=\left( \sum\limits_{ij} \mathrm{Re}(\rho_{ij})|a_i\rangle \langle a_j| \right)  \otimes \left(  \sum\limits_{kl} \mathrm{Re}(\sigma_{kl})|b_k\rangle \langle b_l| \right) \notag \\
    &=\sum\limits_{ijkl} \mathrm{Re}(\rho_{ij})\mathrm{Re}(\sigma_{kl})|a_ib_k\rangle \langle a_jb_l|. \notag
\end{align} Thus $\bigtriangleup^{AB}(\rho \otimes \sigma)=\bigtriangleup^A(\rho) \otimes \bigtriangleup^B(\sigma)$ iff $\mathrm{Re}(\rho_{ij}\sigma_{kl})=\mathrm{Re}(\rho_{ij})\mathrm{Re}(\sigma_{kl})$ for any $i,j,k$ and $l$ iff at least one of $\rho$ and $\sigma$ is a real state. Therefore, item (2) is derived.

(3) Note that
\begin{align}
    &(\bigtriangleup^{A} \otimes \mathbb{I}^B)(\rho \otimes \sigma)\notag \\
    &=\bigtriangleup^{A}\left(\sum\limits_{ij} \rho_{ij}|a_i\rangle \langle a_j| \right)  \otimes \left( \sum\limits_{kl} \sigma_{kl}|b_k\rangle \langle b_l|\right) \notag \\
    &=\sum\limits_{ijkl} \mathrm{Re}\left(\rho_{ij}\right)\sigma_{kl}|a_ib_k\rangle \langle a_jb_l|. \notag
\end{align} So $(\bigtriangleup^{A} \otimes \mathbb{I}^B)(\rho \otimes \sigma)=\bigtriangleup^{AB}(\rho \otimes \sigma)$ iff $\sum\limits_{ijkl} \mathrm{Re}(\rho_{ij})\sigma_{kl}|a_ib_k\rangle \langle a_jb_l|=\sum\limits_{ijkl} \mathrm{Re}(\rho_{ij}\sigma_{kl})|a_ib_k\rangle \langle a_jb_l|$ for any $i,j,k$ and $l$, iff $\sigma$ is a real state. Hence item (3) holds. $\hfill\qedsymbol$ \\\hspace*{\fill}\\
\indent Based on the properties of $\bigtriangleup$, we can use the norm to define imaginarity measures of quantum operations. Now, we define three imaginarity quantifiers of quantum operations as
\begin{equation} \label{eq15}
    M_{c}(\Theta)=\min_{\Phi \in \mathcal{FO_{AB}}} \lVert \Theta \bigtriangleup-\Phi \bigtriangleup\rVert,
\end{equation}
\begin{equation} \label{eq16}
    M_{d}(\Theta)=\min_{\Phi \in \mathcal{FO_{AB}}} \lVert  \bigtriangleup \Theta-\bigtriangleup \Phi\rVert
\end{equation} and
\begin{align} \label{eq17}
    M_{dc}(\Theta)= \lVert  \bigtriangleup \Theta-\Theta
    \bigtriangleup\rVert,
\end{align}
where $\lVert \cdot \rVert$ is any submultiplicative norm satisfying $\lVert \Phi \rVert \le 1$ for any $\Phi \in \mathcal{FO_{AB}}$ and $\lVert \bigtriangleup \rVert \le 1$. Eqs. (\ref{eq15}) and (\ref{eq16}) characterize the ability of quantum operation $\Theta$ to create imaginarity and detect imaginarity, respectively, while Eq. (\ref{eq17}) describes the ability of a quantum operation $\Theta$ to detect and create imaginarity.

\indent {\bf Remark 1} It is worth pointing out that the conditions
for the norms in Eqs. (\ref{eq15}-\ref{eq17}) are not overly
restrictive, since Schatten $p$-norm, a class of commonly used
norms, satisfies these conditions. First of all, it follows from
\cite{WATJ} that the Schatten $p$-norm is submultiplicative. Then
for any $\Theta \in\mathcal{O_{AB}}$, suppose that the eigenvalues
of $\Theta(\rho)$ are $s_{\rho}(0), s_{\rho}(2), \cdots,
s_{\rho}(|B|-1)$. Thus,
\begin{align}
\lVert \Theta  \rVert_p&=\max\left\{\lVert \Phi (\rho)  \rVert_p : \rho \in \mathcal{D(H_A)}, \lVert \rho  \rVert_p \le 1\right\} \notag \\
&=\max \left\{ \left( \mathrm{tr}(\Theta^{\dagger}(\rho) \Theta(\rho))^{\frac{p}{2}}\right) ^{\frac{1}{p}}: \rho \in \mathcal{D(H_A)}, \lVert \rho  \rVert_p \le 1\right\} \notag \\
&=\max\left\{\left(\sum_{i=0}^{|B|-1} \left( s_{\rho}(i)\right)^p \right)^{\frac{1}{p}}: \rho \in \mathcal{D(H_A)}, \lVert \rho  \rVert_p \le 1\right\} \notag \\
&\le\max\left\{\sum_{i=0}^{|B|-1} s_{\rho}(i): \rho \in \mathcal{D(H_A)}, \lVert \rho  \rVert_p \le 1\right\} \notag \\
&\le1, \notag
\end{align} where the first inequality follows from the fact that $\sum\limits_{n}a_n^p \le \left( \sum\limits_{n} a_n\right)^p$ for any $a_n \ge0$ and $p\ge 1$, and the second inequality holds by noting that $\Theta$ is a completely positive map and
trace-nonincreasing. Therefore, $\lVert \Theta\rVert_p\le1$ for any
$\Theta \in \mathcal{O_{AB}}$, which implies that $\lVert
\Phi\rVert_p\le1$ for any $\Phi \in \mathcal{FO_{AB}}$.

Moreover, it holds that
\begin{align}
    \lVert \bigtriangleup \rVert_p &=\max\left\{ \lVert \bigtriangleup(\rho)\rVert_p: \rho \in \mathcal{D(H_A)}, \lVert \rho \rVert_p \le 1 \right\} \notag \\
    &=\max\left\{ \frac{1}{2}\lVert \rho +\rho^{\mathrm{T}}\rVert_p: \rho \in \mathcal{D(H_A)}, \lVert \rho \rVert_p \le 1 \right\} \notag \\
    &\le\max\left\{ \frac{1}{2}(\lVert \rho  \rVert_p+\lVert \rho^{\mathrm{T}}  \rVert_p): \rho \in \mathcal{D(H_A)}, \lVert \rho \rVert_p \le 1\right \} \notag \\
    &\le1, \notag
\end{align} where the last inequality is true because $\lVert \rho  \rVert_p=\lVert \rho^{\mathrm{T}}  \rVert_p \le 1$.

\indent {\bf Theorem 5} $M_c$ defined by Eq. (\ref{eq15}) is an imaginarity measure of quantum operations.\\
\indent \textit {Proof}. Since $\lVert \cdot \rVert$ is a norm, $M(\Theta)=0 \Leftrightarrow \Theta \bigtriangleup=\Phi \bigtriangleup \Leftrightarrow \Theta=\Phi \in \mathcal{FO_{AB}}$. Therefore, $M(\Theta)$ satisfies (M1).\\
\indent Noting that for any $\Phi$, $\tilde{\Phi} \in \mathcal{FO_{AB}}$, $\Phi \tilde{\Phi}$, $\tilde{\Phi} \Phi \in \mathcal{FO_{AB}}$ and using the submultiplicativity of $\lVert \cdot \rVert$, we have
\begin{align}
    M_c(\Theta \tilde{\Phi})&=\min_{\Phi \in \mathcal{FO_{AB}}} \lVert \Theta \tilde{\Phi} \bigtriangleup-\Phi \bigtriangleup\rVert \notag \\
    &\le \min_{\Phi \in \mathcal{FO_{AB}}} \lVert \Theta \tilde{\Phi} \bigtriangleup-\Phi \tilde{\Phi} \bigtriangleup\rVert \notag \\
    &= \min_{\Phi \in \mathcal{FO_{AB}}} \lVert \Theta \bigtriangleup \tilde{\Phi} \bigtriangleup-\Phi \bigtriangleup \tilde{\Phi} \bigtriangleup\rVert \notag \\
    &= \min_{\Phi \in \mathcal{FO_{AB}}} \lVert (\Theta \bigtriangleup-\Phi \bigtriangleup) \tilde{\Phi} \bigtriangleup\rVert \notag \\
    &\le \min_{\Phi \in \mathcal{FO_{AB}}} \lVert \Theta \bigtriangleup-\Phi \bigtriangleup \rVert \lVert \tilde{\Phi} \bigtriangleup\rVert \notag \\
    &\le \min_{\Phi \in \mathcal{FO_{AB}}} \lVert \Theta \bigtriangleup-\Phi \bigtriangleup \rVert \notag \\
    &=M_c(\Theta), \notag
\end{align}
and
\begin{align}
   M_c(\tilde{\Phi}\Theta)&=\min_{\Phi \in \mathcal{FO_{AB}}}  \lVert \tilde{\Phi} \Theta \bigtriangleup-\Phi \bigtriangleup \rVert \notag \\
   &\le \min_{\Phi \in \mathcal{FO_{AB}}}  \lVert \tilde{\Phi} \Theta \bigtriangleup-\tilde{\Phi} \Phi \bigtriangleup  \rVert \notag \\
   &\le \min_{\Phi \in \mathcal{FO_{AB}}}  \lVert \tilde{\Phi}  \rVert  \lVert \Theta \bigtriangleup- \Phi \bigtriangleup  \rVert \notag \\
   &\le \min_{\Phi \in \mathcal{FO_{AB}}}  \lVert \Theta \bigtriangleup- \Phi \bigtriangleup  \rVert \notag \\
   &=M_c(\Theta). \notag
\end{align} Hence (M2a) and (M2b) hold.

Let $p_n \ge 0$, $\sum\limits_{n}p_n=1$, $\Theta_n \in
\mathcal{O_{AB}}$ and $M_c(\Theta_n)=\left \lVert \Theta_n
\bigtriangleup- \Phi_n \bigtriangleup \right \rVert$, where $\Phi_n
\in \mathcal{FO_{AB}}$. Then we find that
\begin{align}
    &M_c\left( \sum_{n}p_n \Theta_n\right)\notag \\
     &=\min_{\Phi \in \mathcal{FO_{AB}}} \left \lVert \left(\sum_{n}p_n \Theta_n\right)\bigtriangleup- \Phi \bigtriangleup \right \rVert \notag \\
    &\le \left \lVert \sum_{n}p_n (\Theta_n\bigtriangleup- \Phi_n \bigtriangleup) \right \rVert \notag \\
    &\le  \sum_{n}p_n \left \lVert \Theta_n\bigtriangleup- \Phi_n \bigtriangleup \right \rVert \notag \\
    &=\sum_{n}p_n M_c(\Theta_n), \notag
\end{align} which implies that (M3) holds. $\hfill\qedsymbol$ \\\hspace*{\fill}\\
\indent {\bf Remark 2} (1) If $\lVert \cdot \rVert$ in Eq.
(\ref{eq15}) is submultiplicative under tensor product and
$\max\limits_{\lVert\tau \rVert\le1}\lVert (\Theta
\bigtriangleup-\Phi \bigtriangleup)\tau\rVert$ achieves its maximum
value when $\tau=\rho \otimes \sigma$, in which at least one of
$\rho$ and $\sigma$ is a real state, then with the help of Theorem
4, we obtain
\begin{align}
    &M_c(\Theta \otimes \mathbb{I})\notag \\
    &=\min_{\Phi \in \mathcal{FO_{AB}}} \lVert (\Theta \otimes \mathbb{I})\bigtriangleup-\Phi \bigtriangleup\rVert \notag \\
    &=\min_{\Phi \in \mathcal{FO_{AB}}} \max_{\lVert\tau\rVert\le1} \lVert ((\Theta \otimes \mathbb{I})\bigtriangleup-\Phi \bigtriangleup)\tau\rVert \notag \\
    &=\min_{\Phi \in \mathcal{FO_{AB}}} \lVert (\Theta \otimes \mathbb{I})\bigtriangleup(\rho \otimes \sigma)-\Phi \bigtriangleup(\rho \otimes \sigma)\rVert \notag \\
    &=\min_{\Phi \in \mathcal{FO_{AB}}} \lVert (\Theta \otimes \mathbb{I})(\bigtriangleup(\rho) \otimes \bigtriangleup(\sigma))-\Phi (\bigtriangleup(\rho) \otimes \bigtriangleup(\sigma))\rVert \notag \\
    &\le\min_{\Phi_1 \in \mathcal{FO_{AB}}} \lVert \Theta\bigtriangleup(\rho) \otimes \bigtriangleup(\sigma)-(\Phi_1\otimes \mathbb{I}) (\bigtriangleup(\rho) \otimes \bigtriangleup(\sigma))\rVert \notag \\
    &=\min_{\Phi_1 \in \mathcal{FO_{AB}}} \lVert ((\Theta\bigtriangleup)(\rho)-(\Phi_1\bigtriangleup)(\rho))\otimes \bigtriangleup(\sigma)\rVert \notag \\
    &\le\min_{\Phi_1 \in \mathcal{FO_{AB}}} \lVert (\Theta\bigtriangleup)(\rho)-(\Phi_1\bigtriangleup)(\rho)\lVert \rVert\bigtriangleup(\sigma)\rVert \notag \\
    &\le\min_{\Phi_1 \in \mathcal{FO_{AB}}} \lVert (\Theta\bigtriangleup)(\rho)-(\Phi_1\bigtriangleup)(\rho) \lVert  \notag \\
    &\le M_c(\Theta). \notag
\end{align}

(2) If $\lVert \cdot \rVert$ in Eq. (\ref{eq15}) is contractive under quantum operations, i.e., $\lVert \mathcal{E}(\rho)-\mathcal{E}(\sigma) \rVert \le \lVert \rho-\sigma \rVert$, where $\mathcal{E} \in \mathcal{O_{AB}}$ and satisfies $\lVert \tau \rVert \le 1$ for any quantum state $\tau$, then we have
\begin{align}
    &M_{c}(\Theta) \notag \\
    &=\min_{\Phi \in \mathcal{FO_{AB}}} \lVert \Theta \bigtriangleup-\Phi \bigtriangleup\rVert \notag \\
    &=\min_{\Phi \in \mathcal{FO_{AB}}} \max_{\rho}\lVert (\Theta \bigtriangleup-\Phi \bigtriangleup)(\rho)\rVert \notag \\
    &=\min_{\Phi \in \mathcal{FO_{AB}}} \max_{\rho} \lVert \mathrm{tr}_{B}((\Theta \bigtriangleup \otimes \mathbb{I})(\rho \otimes |b_0\rangle \langle b_0|))- \mathrm{tr}_{B}((\Phi \bigtriangleup \otimes \mathbb{I})(\rho \otimes |b_0\rangle \langle b_0|)) \rVert \notag \\
    &\le \min_{\Phi \in \mathcal{FO_{AB}}} \max_{\rho} \lVert (\Theta \bigtriangleup \otimes \mathbb{I})(\rho \otimes |b_0\rangle \langle b_0|)- (\Phi \bigtriangleup \otimes \mathbb{I})(\rho \otimes |b_0 \rangle \langle b_0|) \rVert \notag \\
    &= \min_{\Phi \in \mathcal{FO_{AB}}} \max_{\rho} \lVert (\Theta \otimes \mathbb{I})(\bigtriangleup \otimes \mathbb{I})(\rho \otimes |b_0 \rangle \langle b_0|)- (\Phi \otimes \mathbb{I})(\bigtriangleup \otimes \mathbb{I})(\rho \otimes |b_0 \rangle \langle b_0|) \rVert \notag \\
    &=\min_{\Phi \in \mathcal{FO_{AB}}} \max_{\rho} \lVert (\Theta \otimes \mathbb{I})\bigtriangleup (\rho \otimes |b_0\rangle \langle b_0|)- (\Phi \otimes \mathbb{I})\bigtriangleup (\rho \otimes |b_0 \rangle \langle b_0|) \rVert \notag \\
    &\le\min_{\Phi \in \mathcal{FO_{AB}}} \max_{\sigma} \lVert (\Theta \otimes \mathbb{I})\bigtriangleup (\sigma)- (\Phi \otimes \mathbb{I})\bigtriangleup(\sigma) \rVert \notag \\
    &=M_c(\Theta \otimes \mathbb{I}). \notag
\end{align} So $M_{c}(\Theta)\le M_c(\Theta \otimes \mathbb{I})$.

(3) Utilizing the same method, it can be proven that $M_d$  defined by Eq. (\ref{eq16}) is an imaginarity measure of quantum operations. In addition, $M_d$ has the similar property of $M_c$ in item (1), but with the stricter requirement that  $\sigma$ is a real state. And it also satisfies same property of $M_c$ in item (2).

\indent {\bf Theorem 6} $M_{dc}$ defined by Eq. (\ref{eq17}) is an imaginarity measure of quantum operations.\\
\indent \textit {Proof}. We firstly show that $M_{dc}$ satisfies (M1). Note that $M_{dc}(\Theta)=0 \Leftrightarrow \bigtriangleup \Theta= \Theta \bigtriangleup$, i.e., $\Theta \in \mathcal{FO_{AB}}$. Consequently, $M_{dc}$ satisfies (M1).

For any $\Theta\in\mathcal{O_{AB}}$ and $\Phi \in \mathcal{FO_{AB}}$, we have
\begin{align}
    M_{dc}(\Theta \Phi)&=\lVert \bigtriangleup \Theta \Phi- \Theta \Phi \bigtriangleup \rVert  \notag \\
    &=\lVert \bigtriangleup \Theta \Phi- \Theta  \bigtriangleup \Phi \rVert  \notag \\
    &\le \lVert \bigtriangleup \Theta- \Theta \bigtriangleup \rVert \lVert \Phi \rVert \notag \\
    &\le  \lVert \bigtriangleup \Theta- \Theta \bigtriangleup \rVert \notag \\
    &=M_{dc}(\Theta), \notag
\end{align} demonstrating that $M_{dc}$ satisfies (M2a). Similar arguments show that (M2b) holds.

Let $\{\Theta_i\}$ be a set of quantum operations and $p_i \ge 0$ with $\sum\limits_{i} p_i=1$. Then it follows that
\begin{align}
    &M_{dc}\left( \sum\limits_{i} p_i \Theta_i \right)\notag \\
    &=\left \lVert \bigtriangleup \left( \sum\limits_{i} p_i \Theta_i \right) -\left( \sum\limits_{i} p_i \Theta_i \right) \bigtriangleup \right\rVert \notag \\
    &=\left\lVert \sum\limits_{i} p_i \left(\bigtriangleup \Theta_i \right) -\sum\limits_{i} p_i\left(\Theta_i \bigtriangleup \right) \right\rVert \notag \\
    &=\left\lVert \sum\limits_{i} p_i\left( \bigtriangleup \Theta_i -\Theta_i \bigtriangleup \right) \right\rVert \notag \\
    &\le \sum\limits_{i} p_i \lVert \bigtriangleup \Theta_i -\Theta_i \bigtriangleup \rVert \notag \\
    &=\sum\limits_{i} p_i M_{dc}(\Theta_i), \notag
\end{align} which means that (M3) is true for $M_{dc}$. $\hfill\qedsymbol$ \\\hspace*{\fill}

\indent {\bf Remark 3} (1) If $\lVert \cdot \rVert$
in Eq. (\ref{eq17}) is submultiplicative under tensor product and
$\max\limits_{\lVert\tau \rVert\le1}\lVert (\bigtriangleup\Theta
-\Theta \bigtriangleup)\tau\rVert$ achieves its maximum value when
$\tau=\rho \otimes \sigma$, in which $\rho$ is any quantum state and
$\sigma$ is a real state, one finds that
\begin{align}
     &M_{dc}(\Theta \otimes \mathbb{I})\notag\\
     &= \lVert \bigtriangleup(\Theta \otimes \mathbb{I})-(\Theta \otimes \mathbb{I}) \bigtriangleup\rVert \notag \\
    &=\max_{\lVert\tau\rVert\le1} \lVert \left(\bigtriangleup(\Theta \otimes \mathbb{I})-(\Theta \otimes \mathbb{I}) \bigtriangleup \right) \left( \tau\right) \rVert \notag \\
    &=\lVert \bigtriangleup(\Theta \otimes \mathbb{I})(\rho \otimes \sigma)-(\Theta \otimes \mathbb{I}) \bigtriangleup(\rho \otimes \sigma)\rVert \notag \\
    &=\lVert (\bigtriangleup\Theta) (\rho) \otimes \bigtriangleup(\sigma)-(\Theta \otimes \mathbb{I}) (\bigtriangleup(\rho) \otimes \bigtriangleup(\sigma))\rVert \notag \\
    &=\lVert (\bigtriangleup\Theta) (\rho) \otimes \bigtriangleup(\sigma)-(\Theta\bigtriangleup)(\rho) \otimes \bigtriangleup(\sigma) \rVert \notag \\
    &=\lVert \left( (\bigtriangleup\Theta) (\rho) -(\Theta\bigtriangleup)(\rho)\right)  \otimes \bigtriangleup(\sigma) \rVert \notag \\
    &=\lVert (\bigtriangleup\Theta) (\rho) -(\Theta\bigtriangleup)(\rho) \rVert \rVert\bigtriangleup(\sigma)\rVert \notag \\
    &=\max_{\lVert\rho\rVert\le1} \lVert (\bigtriangleup\Theta) (\rho) -(\Theta\bigtriangleup)(\rho) \rVert \notag \\
    &\le M_{dc}(\Theta), \notag
\end{align} where the fourth and fifth equality holds by using items (2) and (3) of Theorem 4, respectively, the seventh equality follows from the fact that $\lVert\cdot\rVert$ is submultiplicative under tensor product and the eighth equality is true because $\lVert\bigtriangleup(\sigma)\rVert\le\lVert\bigtriangleup \rVert \lVert \sigma \rVert \le 1$. In addition, similar arguments show that Remark 2 (2) also holds for $M_{dc}$.

(2) If there exists $\Phi \in \mathcal{FO_{AB}}$ such
that $\bigtriangleup\Theta=[(1+k)\Phi-k\Theta]\bigtriangleup$ for
any $k\in \mathbb{R}$, with the help of the triangle inequality and
homogeneity of norm, the relations among $M_c(\Theta)$,
$M_d(\Theta)$ and $M_{dc}(\Theta)$ can be characterized as
\begin{align}\label{eq18}
        M_{dc}(\Theta)&=\lVert  \bigtriangleup \Theta-\Theta \bigtriangleup\rVert \notag \\
        &=\lVert \bigtriangleup \Theta-\bigtriangleup \Phi \rVert+\lVert \Theta \bigtriangleup-\Phi \bigtriangleup\rVert  \notag \\
        &= \min_{\Phi \in \mathcal{FO_{AB}}} \left\{\lVert \bigtriangleup \Theta-\bigtriangleup \Phi \rVert+\lVert \Theta \bigtriangleup-\Phi \bigtriangleup\rVert  \right\} \notag \\
    &\ge \min_{\Phi \in \mathcal{FO_{AB}}}\lVert \bigtriangleup \Theta-\bigtriangleup \Phi \rVert+\min_{\Phi \in \mathcal{FO_{AB}}} \lVert \Theta \bigtriangleup-\Phi \bigtriangleup\rVert  \notag \\
    &=M_c(\Theta) + M_d(\Theta),
\end{align} where the second equality is true because $\bigtriangleup \Theta-\bigtriangleup \Phi=-k(\Theta \bigtriangleup-\Phi \bigtriangleup)$.

\indent {\bf Theorem 7} $M_{dc}$ defined by Eq. (\ref{eq17}) attains its maximum value at a pure state.\\
\indent \textit {Proof}. For any mixed state $\rho$, we can
spectrally decompose it as $\rho=\sum\limits_{i} p_i |\psi_i \rangle
\langle \psi_i|$ where $|\psi_i \rangle$ are eigenstates of $\rho$. We denote the set of pure states on $\mathcal{H_A}$ by $\mathcal{PS_A}$.
Utilizing the convexity of $M_{dc}(\Theta)$, one finds that
\begin{align}
    &\lVert  (\bigtriangleup \Theta-\Theta \bigtriangleup)\rho\rVert \notag \\
    &= \left\lVert  (\bigtriangleup \Theta-\Theta \bigtriangleup)\left(\sum\limits_{i} p_i |\psi_i \rangle \langle \psi_i| \right) \right\rVert \notag \\
    &=\left\lVert \sum\limits_{i} p_i (\bigtriangleup \Theta-\Theta \bigtriangleup) |\psi_i \rangle \langle \psi_i|  \right\rVert \notag \\
    &\le \sum_{i}p_i \left\lVert (\bigtriangleup \Theta-\Theta \bigtriangleup) |\psi_i \rangle \langle \psi_i|  \right\rVert \notag \\
    &\le \max_{|\psi \rangle\in \mathcal{PS_A}}\left\lVert (\bigtriangleup \Theta-\Theta \bigtriangleup) |\psi \rangle \langle \psi|  \right\rVert. \notag
\end{align} Hence,
\begin{align}
        M_{dc}(\Theta)&= \lVert  \bigtriangleup \Theta-\Theta \bigtriangleup\rVert \notag\\&=\max_{\rho\in\mathcal{D(H_A)}}\lVert  (\bigtriangleup \Theta-\Theta \bigtriangleup)\rho\rVert \notag \\
        &\le \max_{|\psi \rangle\in \mathcal{PS_A}}\left\lVert (\bigtriangleup \Theta-\Theta \bigtriangleup) |\psi \rangle \langle \psi|  \right\rVert. \notag
\end{align}
On the other hand, the reverse inequality holds obviously.
Therefore,
\begin{align}
        M_{dc}(\Theta)=\max_{|\psi \rangle\in \mathcal{PS_A}}\left\lVert (\bigtriangleup \Theta-\Theta \bigtriangleup) |\psi \rangle \langle \psi|  \right\rVert.
        \notag
\end{align} $\hfill\qedsymbol$ \\\hspace*{\fill}\\
 \indent Note that the trace norm satisfies all the conditions for the norm in Eq. (\ref{eq17}). Taking the norm as the trace norm,
 we get the following quantifier
 \begin{align}\label{eq19}
 M_{dc}^t(\Theta)= \lVert  \bigtriangleup \Theta-\Theta \bigtriangleup\rVert_1,
 \end{align} where $\lVert A \rVert_1=\mathrm{tr}\sqrt{A^{\dagger}A}$. Using triangle inequality of the norm, we find that
 \begin{align}
     &\lVert  \bigtriangleup \Theta-\Theta \bigtriangleup\rVert_1 \notag \\
     &=\max \left\{ \lVert ( \bigtriangleup \Theta-\Theta \bigtriangleup)\rho\rVert_1: \rho \in \mathcal{D(H_A)}, \lVert \rho \rVert_1 \le1 \right \} \notag \\
     &\le \max_{\rho \in \mathcal{D(H_A)}} \left\{ \lVert  \bigtriangleup \Theta\rho\rVert_1+\lVert \Theta \bigtriangleup\rho \rVert_1 \right \} \notag \\
     &\le \max_{\rho \in \mathcal{D(H_A)}}  \lVert  \bigtriangleup \Theta\rho\rVert_1 + \max_{\rho \in \mathcal{D(H_A)}}  \lVert  \Theta\bigtriangleup \rho\rVert_1 \notag \\
     &\le \max_{\rho \in \mathcal{D(H_A)}} \lVert   \Theta\rho\rVert_1 + \max_{\rho \in \mathcal{D(H_A)}}  \lVert   \Theta\rho\rVert_1 \notag \\
     &\le 2, \notag
 \end{align}  where the last inequality follows from the fact that quantum operations are trace-nonincreasing. Namely, an upper bound of $M_{dc}^t(\Theta)$ is 2.

\indent The weights of coherence of quantum states\cite{BKAN} and
quantum channels\cite{LYYM} have been introduced and studied, while
the weight of imaginarity of quantum states\cite{XSGJ} have also
been proposed, which bear significant physical meaning and find
important physical applications. We define the
weight of imaginarity of quantum operations as
\begin{align}\label{eq20}
    \mathscr{I}_w(\Theta)=\min\left \{ 0\le s\le1: \Theta=(1-s) \Phi +s\Lambda, {\rm{\ for \ some\ }} \Phi \in \mathcal{FO_{AB}} {\rm{\ and\ }} \Lambda \in \mathcal{O_{AB}} \right \}.
\end{align}

Now we demonstrate that the above definition is well-defined. First
of all, for any given $\Theta$, denote by $S_{\Theta}$ the set of
all $s$ satisfying $\Theta=(1-s) \Phi +s\Lambda$, for some $\Phi \in
\mathcal{FO_{AB}}$ and $\Lambda \in \mathcal{O_{AB}}$. It is obvious
that $0\in S_{\Theta}$ if $\Theta\in\mathcal{FO_{AB}}$, while $1\in
S_{\Theta}$ if $\Theta\notin\mathcal{FO_{AB}}$, which implies that
$S_{\Theta}$ is nonempty. Next, we show that $S_{\Theta}$ is closed.
In fact, for any convergent sequence $\{s_n\}\subseteq S_{\Theta}$,
suppose that $s_n\rightarrow s_0$ when $n\rightarrow\infty$. Since
$s_n\in S_{\Theta}$, there exist $\Phi_n\in\mathcal{FO_{AB}}$ and
$\Lambda_n\in\mathcal{O_{AB}}$ such that
\begin{align}
    \Theta=(1-s_n)\Phi_n+s_n\Lambda_n \notag
\end{align} for each $n$. Since $\mathcal{FO_{AB}}$ and $\mathcal{O_{AB}}$ are compact sets\cite{WATJ}, $\{\Phi_n\}\subseteq \mathcal{FO_{AB}}$ and $\{\Lambda_n\}\subseteq \mathcal{O_{AB}}$, there exist convergent subsequences $\{\Phi_{n_k}\}$ and $\{\Lambda_{n_k}\}$ such that
\begin{align}
    \Phi_{n_k}\rightarrow \Phi^* {\rm{\ and\ }} \Lambda_{n_k}\rightarrow \Lambda^*, \notag
\end{align} when $k\rightarrow\infty$, respectively. Noting that both of $\mathcal{FO_{AB}}$ and $\mathcal{O_{AB}}$ are closed, we have $\Phi^*\in\mathcal{FO_{AB}}$ and $\Lambda^*\in\mathcal{O_{AB}}$.
Thus
\begin{align}
    \Theta&=\lim_{k\rightarrow\infty}\left[(1-s_{n_k})\Phi_{n_k}+s_{n_k}\Lambda_{n_k}\right] \notag\\
    &=(1-s_0)\Phi^*+s_0\Lambda^*, \notag
\end{align} which indicates that $s_0\in S_{\Theta}$. Therefore, $S_{\Theta}$ is a closed set. Since $S_{\Theta}$ is nonempty and closed, the minimum in Eq. (\ref{eq20}) is attainable.

\indent {\bf Theorem 8} $\mathscr{I}_w$ defined by Eq. (\ref{eq20}) satisfies the following properties: \\
\indent (1) $\mathscr{I}_w (\Theta_1 \circ \Theta_2) \le \mathscr{I}_w (\Theta_1)+\mathscr{I}_w (\Theta_2)-\mathscr{I}_w (\Theta_1)\mathscr{I}_w (\Theta_2)$ for any $\Theta_1\in\mathcal{O_{AB}}$ and $\Theta_2\in\mathcal{O_{CA}}$.\\
\indent (2) $\mathscr{I}_w (\Theta_1 \otimes \Theta_2) \le \mathscr{I}_w (\Theta_1)+\mathscr{I}_w (\Theta_2)-\mathscr{I}_w (\Theta_1)\mathscr{I}_w (\Theta_2)$ for any $\Theta_1\in\mathcal{O_{AB}}$ and $\Theta_2\in\mathcal{O_{CD}}$.\\
\indent \textit {Proof}. (1) Suppose that
\begin{align}
    (1-\mathscr{I}_w (\Theta_1))\Phi_1+\mathscr{I}_w(\Theta_1)\Lambda_1 \notag
\end{align} and
\begin{align}
    (1-\mathscr{I}_w (\Theta_2))\Phi_2+\mathscr{I}_w(\Theta_2)\Lambda_2 \notag
\end{align} are the optimal decompositions of $\Theta_1$ and $\Theta_2$, respectively, where $\Phi_1\in\mathcal{FO_{AB}}$, $\Lambda_1\in\mathcal{O_{AB}}$, $\Phi_2\in\mathcal{FO_{CA}}$ and $\Lambda_2\in\mathcal{O_{CA}}$. Then we obtain
\begin{align}
    \Theta_1 \circ \Theta_2&=(1-\mathscr{I}_w (\Theta_1))(1-\mathscr{I}_w (\Theta_2)) \Phi_1 \circ \Phi_2+(1-\mathscr{I}_w (\Theta_1))\mathscr{I}_w (\Theta_2) \Phi_1 \circ \Lambda_2 \notag \\
    &+\mathscr{I}_w (\Theta_1)(1-\mathscr{I}_w (\Theta_2)) \Lambda_1 \circ \Phi_2 +\mathscr{I}_w (\Theta_1)\mathscr{I}_w (\Theta_2) \Lambda_1 \circ \Lambda_2. \notag
\end{align}
Let $s_t=\mathscr{I}_w (\Theta_1)+\mathscr{I}_w
(\Theta_2)-\mathscr{I}_w (\Theta_1)\mathscr{I}_w (\Theta_2)$. It is
obvious that $0\le s_t\le1$.

\textbf{Case 1} $s_t=0$.

In this case, $\mathscr{I}_w(\Theta_1)=\mathscr{I}_w(\Theta_2)=0$. It follows from Eq. (\ref{eq20}) that $\Theta_1$ and $\Theta_2\in \mathcal{FO_{AB}}$. Therefore, one has $\Theta_1\circ\Theta_2\in\mathcal{FO_{CB}}$, which implies that $\mathscr{I}_w(\Theta_1\circ\Theta_2)=0=\mathscr{I}_w (\Theta_1)+\mathscr{I}_w (\Theta_2)-\mathscr{I}_w (\Theta_1)\mathscr{I}_w (\Theta_2)$.

\textbf{Case 2} $0<s_t\le1$.

Letting $\Phi_t=\Phi_1 \circ \Phi_2$ and $\Lambda_t=\frac{1}{s_t} [(1-\mathscr{I}_w (\Theta_1))\mathscr{I}_w (\Theta_2) \Phi_1 \circ \Lambda_2 +\mathscr{I}_w (\Theta_1)(1-\mathscr{I}_w (\Theta_2)) \Lambda_1 \circ \Phi_2 +\mathscr{I}_w (\Theta_1)\mathscr{I}_w (\Theta_2) \Lambda_1 \circ \Lambda_2 ]$, we get $\Theta=(1-s_t)\Phi_t+s_t \Lambda_t$. It follows that $\mathscr{I}_w(\Theta_1 \circ \Theta_2) \le s_t=\mathscr{I}_w (\Theta_1)+\mathscr{I}_w (\Theta_2)-\mathscr{I}_w (\Theta_1)\mathscr{I}_w (\Theta_2)$. Therefore, item (1) is derived.

(2) First of all, we claim that
$\Phi_1\otimes\Phi_2\in\mathcal{FO_{ACBD}}$ for any
$\Phi_1\in\mathcal{FO_{AB}}$ and $\Phi_2\in\mathcal{FO_{CD}}$. In
fact, let
\begin{align}
   \Phi_1(|a_i\rangle \langle a_j|)=\sum_{k,l}(\Phi_1)_{k,l}^{i,j}|b_k\rangle \langle b_l| \notag
\end{align} and
\begin{align}
     \Phi_2(|c_m\rangle \langle c_n|)=\sum_{k,l}(\Phi_2)_{s,t}^{m,n}|d_s\rangle \langle d_t|.\notag
\end{align} Then we have
\begin{align}
    &(\Phi_1\otimes\Phi_2)(|a_i c_m \rangle \langle  a_j c_n|) \notag \\
    &=  \left(\sum_{k,l}(\Phi_1)_{k,l}^{i,j}|b_k\rangle \langle b_l| \right)\otimes  \left( \sum_{k,l}(\Phi_2)_{s,t}^{m,n}|d_s\rangle \langle d_t| \right)\notag\\
    &=\sum_{k,l,s,t} (\Phi_1)_{k,l}^{i,j}(\Phi_2)_{s,t}^{m,n}|b_k d_s    \rangle \langle b_l d_t |.\notag
\end{align} Since $\Phi_1\in\mathcal{FO_{AB}}$ and $\Phi_2\in\mathcal{FO_{CD}}$, it follows from Theorem 2 that $(\Phi_1)_{k,l}^{i,j}$, $(\Phi_2)_{s,t}^{m,n}\in\mathbb{R}$, and thus $(\Phi_1)_{k,l}^{i,j}(\Phi_2)_{s,t}^{m,n} \in\mathbb{R}$ for any $i,j,k,l,m,n,s$ and $t$. Utilizing Theorem 2 again, we obtain $\Phi_1\otimes\Phi_2\in\mathcal{FO_{ACBD}}$. Using the method of item (1), we can then prove item (2). $\hfill\qedsymbol$ \\\hspace*{\fill}\\
\indent It can be seen from Theorem 8 that $\mathscr{I}_w (\Theta_1 \circ \tilde{\Phi}) \le \mathscr{I}_w (\Theta_1)$ and $\mathscr{I}_w (\tilde{\Phi} \circ \Theta_2) \le \mathscr{I}_w (\Theta_2)$ for any free operation $\tilde{\Phi}$. Utilizing this fact, we can prove the following theorem.\\
\indent {\bf Theorem 9} $\mathscr{I}_w$ defined by Eq. (\ref{eq20}) is an imaginarity measure of quantum operations.\\
\indent \textit {Proof}. Note that $\mathscr{I}_w(\Theta)=0
\Leftrightarrow \Theta =\Phi \in \mathcal{FO_{AB}}$.
$\mathscr{I}_w(\Theta)$ thus satisfies (M1). Meanwhile, it follows
from Theorem 8 that (M2a) and (M2b) both holds for
$\mathscr{I}_w(\Theta)$.

Then we prove that $\mathscr{I}$ satisfies (M3). Let $\{\Theta_i\}$ be a set of quantum operations from $\mathcal D( \mathcal {H_A})$ to $\mathcal D( \mathcal {H_B})$, $p_i\ge0$ with $\sum\limits_{i}p_i=1$, $\hat{s}=\sum_i p_i \mathscr{I}_w (\Theta_i)$ and $\mathcal{I^+}=\{i:p_i>0\}$. \\
\indent \textbf{Case 1} $\hat{s}=0$.\\
\indent It is obvious that $\mathscr{I}_w(\Theta_i)=0$ for $i\in \mathcal{I^+}$, i.e., $\Theta_i\in\mathcal{FO_{AB}}$ for $i\in \mathcal{I^+}$. We thus have $\mathscr{I}_w (\sum_i p_i \Theta_i)= 0=\sum_i p_i \mathscr{I}_w (\Theta_i)$.\\
\indent \textbf{Case 2} $\hat{s}=1$.\\
\indent First of all, we claim that $\mathscr{I}_w(\sum_i p_i\Theta_i)=1$ if $\mathscr{I}_w(\Theta_i)=1$ for $i\in \mathcal{I^+}$. In fact, suppose that $\mathscr{I}_w(\sum_ip_i\Theta_i)\neq1$, which implies that there exist $\Phi\in\mathcal{FO_{AB}}$ and $\Lambda\in\mathcal{O_{AB}}$ such that $\sum_i p_i\Theta_i=(1-s)\Phi+s\Lambda=\sum_i p_i\left((1-s)\Phi+s\Lambda \right)$, where $0\leq s<1$. This implies that $\Theta_i=(1-s)\Phi+s\Lambda$ for $i\in \mathcal{I^+}$, and thus $\mathscr{I}_w(\Theta_i)\leq s<1$ for $i\in \mathcal{I^+}$, which is a contradiction. So this claim is true.\\
\indent Since $\hat{s}=1$, we have $\mathscr{I}_w(\Theta_i)=1$ for $i\in \mathcal{I^+}$. By using the above claim, we obtain that $\mathscr{I}_w (\sum_i p_i \Theta_i)=1=\sum_i p_i \mathscr{I}_w (\Theta_i)$.\\
\indent \textbf{Case 3} $0<\hat{s}<1$.\\
\indent Let
\begin{align}
    \Theta_i=(1-\mathscr{I}_w (\Theta_i))\Phi_i+\mathscr{I}_w (\Theta_i) \Lambda_i \notag
\end{align} be the optimal decompositions of $\Theta_i$. Then we have
\begin{align}
    \sum_i p_i \Theta_i =\sum_i p_i (1-\mathscr{I}_w (\Theta_i))\Phi_i +\sum_i p_i \mathscr{I}_w (\Theta_i) \Lambda_i. \notag
\end{align} Setting $\hat{\Phi}=\frac{1}{1-\hat{s}}(\sum\limits_i p_i (1-\mathscr{I}_w (\Theta_i))\Phi_i)$ and $\hat{\Lambda}=\frac{1}{\hat{s}}\sum_i p_i \mathscr{I}_w (\Theta_i) \Lambda_i$, we thus get
\begin{align}
        \sum_i p_i \Theta_i = (1-\hat{s}) \hat{\Phi}+ \hat{s} \hat{\Lambda}, \notag
\end{align} which implies that $\mathscr{I}_w (\sum_i p_i \Theta_i) \le \hat{s}=\sum_i p_i \mathscr{I}_w (\Theta_i)$. So (M3) is proved. $\hfill\qedsymbol$ \\\hspace*{\fill}\\

We now turn to discuss the relations between the
imaginarity measures we defined based on the norm and the weight.
Denote by $M_c^p(\Theta)$ the quantity by taking the norm in Eq.
(\ref{eq15}) as the Schatten $p$-norms, i.e.,
$M_c^p(\Theta)=\min\limits_{\Phi \in \mathcal{FO_{AB}}} \lVert \Theta
\bigtriangleup-\Phi \bigtriangleup\rVert_p$. We first have the
following results.

\indent {\bf Theorem 10} It holds that

(1) $M_c^p(\Theta)\le\mathscr{I}_w(\Theta)$ for any
$\Theta\in\mathcal{O_{AB}}$;

(2)
$M_c^p(\Omega\circ\Theta)\le\mathscr{I}_w(\Theta)$ for any
$\Theta\in\mathcal{O_{AB}}$ and $\Omega\in\mathcal{FO_{BC}}$, and
$M_c^p(\Theta\circ\Omega)\le\mathscr{I}_w(\Theta)$ for any
$\Theta\in\mathcal{O_{AB}}$ and $\Omega\in\mathcal{FO_{DA}}$;

(3)
$M_c^p(\Omega\otimes\Theta)\le\mathscr{I}_w(\Theta)$ and
$M_c^p(\Theta\otimes\Omega)\le\mathscr{I}_w(\Theta)$ for any
$\Theta\in\mathcal{O_{AB}}$ and $\Omega\in\mathcal{FO_{CD}}$.

 \textit {Proof}. Suppose that
\begin{align}
    \Theta=\left(1-\mathscr{I}_w(\Theta)\right)\Phi^*+\mathscr{I}_w(\Theta)\Lambda^*\notag
\end{align} is the optimal decomposition of $\Theta$, where $\Phi^*\in\mathcal{FO_{AB}}$ and $\Lambda^*\in\mathcal{O_{AB}}$.

(1) Note that
\begin{align}
    &M_{c}^p(\Theta)\notag \\
    &=\min_{\Phi \in \mathcal{FO_{AB}}} \lVert \Theta \bigtriangleup-\Phi \bigtriangleup\rVert_p\notag\\
    &\le\min_{\Phi \in \mathcal{FO_{AB}}} \lVert \Theta \bigtriangleup-\Phi \bigtriangleup\rVert_1\notag\\
    &=\min_{\Phi \in \mathcal{FO_{AB}}} \left\lVert \left(\left(1-\mathscr{I}_w(\Theta)\right)\Phi^*+\mathscr{I}_w(\Theta)\Lambda^* \right)  \bigtriangleup-\Phi \bigtriangleup\right\rVert_1\notag\\
    &=\min_{\Phi \in \mathcal{FO_{AB}}} \left\lVert \left(1-\mathscr{I}_w(\Theta)\right)\Phi^* \bigtriangleup-\Phi \bigtriangleup+\mathscr{I}_w(\Theta)\Lambda^*\bigtriangleup\right\rVert_1\notag\\
    &\le\tilde{\mathscr{I}_w}(\Theta)\left\lVert\Lambda^*\bigtriangleup\right\rVert_1\notag\\
    &\le\mathscr{I}_w(\Theta)\left\lVert\Lambda^*\right\rVert_1\notag\\
    &\le\mathscr{I}_w(\Theta), \notag
\end{align} where the first inequality follows from the fact that for any operator $A$ and $1\le p\le q<+\infty$, $\lVert A\rVert_p \ge \lVert A\rVert_q$ holds, the second inequality is true by taking $\Phi=\left(1-\mathscr{I}_w(\Theta)\right)\Phi^*$, the third inequality holds since $\lVert\cdot\rVert_1$ is submultiplicative and $\lVert\bigtriangleup\rVert_1 \le1$, and the last inequality follows because $\Lambda^*$ is trace-nonincreasing. So item (1) is true.

(2) Since
$\mathcal{E}_{1,\Omega}(\Theta)=\Omega\circ\Theta$ and
$\mathcal{E}_{2,\Omega}(\Theta)=\Theta\circ\Omega$ are both free
superoperations for any $\Omega\in\mathcal{FO_{AB}}$ and
$M_c^p(\Theta)$ is a bona fide imaginarity measure, we have
\begin{align}
    M_c^p(\Omega\circ\Theta)\le M_c^p(\Theta) \notag
\end{align}
and
\begin{align}
    M_c^p(\Theta\circ\Omega)\le M_c^p(\Theta). \notag
\end{align}
Combining these with item (1), item (2) follows immediately.

(3) From the proof of the item (2) in Theorem 8, we know that $\Omega\otimes\Phi^*\in\mathcal{FO_{CADB}}$ for any
$\Phi^*\in\mathcal{FO_{AB}}$ and $\Omega\in\mathcal{FO_{CD}}$. Then direct calculations show that
\begin{align}
        &M_{c}^p(\Omega\otimes \Theta)\notag \\
        &=\min_{\Phi \in \mathcal{FO_{AB}}} \lVert (\Omega\otimes \Theta) \bigtriangleup-\Phi \bigtriangleup\rVert_p\notag\\
        &=\min_{\Phi \in \mathcal{FO_{AB}}} \lVert \Omega\otimes \left[\left(1-\mathscr{I}_w(\Theta)\right)\Phi^*+\mathscr{I}_w(\Theta)\Lambda^*\right] \bigtriangleup-\Phi \bigtriangleup\rVert_p\notag\\
        &\le\min_{\Phi \in \mathcal{FO_{AB}}} \lVert \Omega\otimes \left[\left(1-\mathscr{I}_w(\Theta)\right)\Phi^*+\mathscr{I}_w(\Theta)\Lambda^*\right] \bigtriangleup-\Phi \bigtriangleup\rVert_1\notag\\
        &=\min_{\Phi \in \mathcal{FO_{AB}}} \lVert \left(1-\mathscr{I}_w(\Theta)\right)(\Omega\otimes \Phi^*)\bigtriangleup+\mathscr{I}_w(\Theta)(\Omega\otimes\Lambda^* ) \bigtriangleup-\Phi \bigtriangleup\rVert_1\notag\\
        &\le\lVert \mathscr{I}_w(\Theta)(\Omega\otimes\Lambda^* ) \bigtriangleup\rVert_1\notag\\
        &\le\mathscr{I}_w(\Theta)\lVert\Omega\otimes\Lambda^*\rVert_1\notag\\
        &\le\mathscr{I}_w(\Theta), \notag
\end{align} where the first inequality follows from the property of the Schatten $p$ norms,
the second inequality holds by taking
$\Phi=\left(1-\mathscr{I}_w(\Theta)\right)\Omega\otimes \Phi^*$, in
which $\Omega \otimes\Phi^*$ is a free operation from the above
claim, the third inequality holds since $\lVert\cdot\rVert_1$ is
submultiplicative and $\lVert\bigtriangleup\rVert_1\le 1$, and the
last inequality follows because $\Omega\otimes\Lambda^*$ is
trace-nonincreasing. In a similar way, we can verify that
$M_c^p(\Theta\otimes\Omega)\le\mathscr{I}_w(\Theta)$. Hence item (3)
holds.$\hfill\qedsymbol$

\indent {\bf Remark 4} Denote by $M_d^p(\Theta)$ the
quantity by taking the norm in Eq. (\ref{eq16}) as the Schatten
$p$-norms. It can be seen that the properties in Theorem 10 also
hold for $M_d^p(\Theta)$.

More generally, we clarify the relations between
$\mathscr{I}_w$ and $M_{dc}$ as follows.

\indent {\bf Theorem 11} Suppose that $\Theta=(1-s^*)\Phi^*+s^* \Lambda^*$ is the optimal decomposition of $\Theta$. Then we have \\
\indent (1) $M_{dc}(\Theta)=\mathscr{I}_w (\Theta)M_{dc}(\Lambda^*)$ for any $\Theta \in \mathcal{O_{AB}}$.\\
\indent (2) $M_{dc}( \Omega\circ \Theta)=\mathscr{I}_w (\Theta)M_{dc}(\Omega\circ\Lambda^*)$ for any $\Theta \in \mathcal{O_{AB}}$ and $\Omega \in \mathcal{FO_{BC}}$, and $M_{dc}(\Theta \circ \Omega)=\mathscr{I}_w (\Theta)M_{dc}(\Lambda^* \circ \Omega)$ for any $\Theta \in \mathcal{O_{AB}}$ and $\Omega \in \mathcal{FO_{DA}}$.\\
\indent (3) $M_{dc}( \Omega\otimes \Theta)=\mathscr{I}_w
(\Theta)M_{dc}(\Omega\otimes \Lambda^*)$ and $M_{dc}(\Theta \otimes
\Omega)=\mathscr{I}_w (\Theta)M_{dc}(\Lambda^* \otimes \Omega)$ for
any $\Theta \in \mathcal{O_{AB}}$ and $\Omega \in
\mathcal{FO_{CD}}$.

 \textit {Proof}. (1) Note that
\begin{align}
    \bigtriangleup \Theta=(1-s^*)\bigtriangleup \Phi^*+s^*\bigtriangleup \Lambda^* \notag
\end{align} and
\begin{align}
    \Theta \bigtriangleup=(1-s^*)\Phi^* \bigtriangleup +s^*\Lambda^* \bigtriangleup.  \notag
\end{align} Since $\Phi \in \mathcal{FO_{AB}}$, we obtain
\begin{align}
    \bigtriangleup \Theta-\Theta \bigtriangleup=s^*(    \bigtriangleup \Lambda^*-\Lambda^* \bigtriangleup), \notag
\end{align} which implies that
\begin{align}
    \lVert \bigtriangleup \Theta-\Theta \bigtriangleup \rVert=s^*\lVert(\bigtriangleup \Lambda^*-\Lambda^* \bigtriangleup)\rVert, \notag
\end{align}where $\lVert \cdot \rVert$ satisfies the condition in Eq. (\ref{eq17}). Therefore, it follows from Eqs. (\ref{eq17}) and (\ref{eq20}) that
\begin{align}
M_{dc}(\Theta)=\mathscr{I}_w (\Theta)M_{dc}(\Lambda^*). \notag
\end{align} Using the same method, we obtain that $M_{dc}( \Omega\circ \Theta)=\mathscr{I}_w (\Theta)M_{dc}(\Omega\circ\Lambda^*)$.

(2) Noting that
\begin{align}
    \bigtriangleup ( \Omega\circ \Theta)- (\Omega \circ\Theta ) \bigtriangleup =s^* \left[\bigtriangleup (\Omega\circ \Lambda^*) -(\Omega\circ\Lambda^* ) \bigtriangleup\right],\notag
\end{align} we obtain
\begin{align}
        \lVert \bigtriangleup ( \Omega\circ \Theta)- (\Omega\circ \Theta)  \bigtriangleup \rVert =s^*\lVert\bigtriangleup (\Omega\circ \Lambda^*) -(\Omega\circ \Lambda^*) \bigtriangleup\rVert, \notag
\end{align} where $\lVert \cdot \rVert$ satisfies the condition in Eq. (\ref{eq17}). Thus, by Eqs. (\ref{eq17}) and (\ref{eq20}), we have
\begin{align}
    M_{dc}(\Omega\circ \Theta)=\mathscr{I}_w (\Theta)M_{dc}(\Omega\circ \Lambda^*)
    \notag.
\end{align} In a similar way, we get $M_{dc}( \Theta\circ\Omega )=\mathscr{I}_w
(\Theta)M_{dc}(\Lambda^*\circ\Omega)$.

(3) Imitating the proof of item (2), we can immediately obtain item
(3).$\hfill\qedsymbol$ \\

\noindent {\bf 5. An example}\\\hspace*{\fill}\\
\indent In this section, we present an example to illustrate that
the explicit formulas of the quantity in Eq. (\ref{eq19}) can be
derived for qubit unitary operations, and the values of the quantity
for unitary operations induced by typical qubit quantum gates can be
calculated.

\indent {\bf Example 1} Consider the quantum operation $\Theta$ in
Eq. (\ref{eq19}) as the one induced by qubit unitary
gates\cite{BABH}
\begin{align}
    U(\theta,\phi,\lambda)= \begin{pmatrix}
        \cos\frac{\theta}{2} & -e^{\mathrm{i}\lambda}\sin\frac{\theta}{2} \\
        e^{\mathrm{i}\phi}\sin\frac{\theta}{2} & e^{\mathrm{i}(\phi+\lambda)}\cos\frac{\theta}{2}
    \end{pmatrix}, \notag
\end{align} where $\theta,\phi,\lambda \in [0, 2\pi]$. Any qubit state $\rho$
can be expressed in Bloch representation as
\begin{equation}
    \rho=\frac{1}{2}(\mathbf{I_2}+ \vec r \cdot \vec \sigma)=\frac{1}{2}
    \begin{pmatrix}
        1+r_{3} & r_{1}-\mathrm{i}r_2\\
        r_1+\mathrm{i}r_2 & 1-r_3\\
    \end{pmatrix}, \notag
\end{equation} where $\mathbf{I_2}$ is the $2\times 2$ identity matrix, $\vec r =(r_1,r_2,r_3)$ is a real vector with $r=|\vec r|\le 1$ and $\vec \sigma=(\sigma_{x},\sigma_{y},\sigma_{z})$ with $\sigma_{x}=
\begin{pmatrix}
    0 & 1\\
    1 & 0\\
\end{pmatrix}$, $\sigma_{y}=
\begin{pmatrix}
    0 & -\mathrm{i}\\
    \mathrm{i} & 0\\
\end{pmatrix}$ and $\sigma_{z}=
\begin{pmatrix}
    1 & 0\\
    0 & -1\\
\end{pmatrix}$.

\begin{figure}[H]\centering
    {\begin{minipage}[figure2]{0.49\linewidth}
            \includegraphics[width=1.0\textwidth]{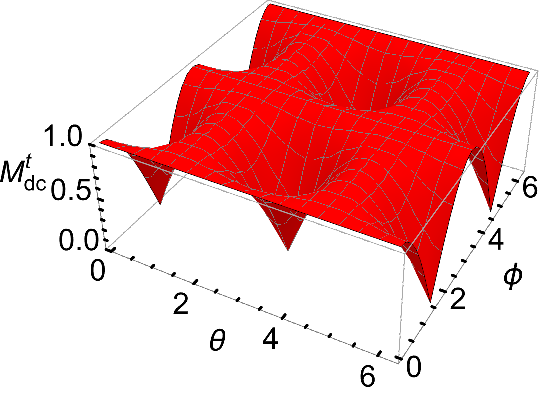}
    \end{minipage}}
    \caption{{The variations of $M_{dc}^t(\hat{\Theta})$ in Eq. (\ref{eq22}) for fixed $\lambda=\frac{\pi}{2}$ \label{fig:Fig2}}}
\end{figure}

By direct calculations, we have
\begin{align}
    (\bigtriangleup \Theta-\Theta \bigtriangleup)\rho=\left(
    \begin{array}{cc}
        \frac{1}{2} r_2 \sin \theta  \sin \lambda  & t \\
        t^* & -\frac{1}{2} r_2 \sin \theta  \sin \lambda  \\
    \end{array}
    \right), \notag
\end{align} where $t=\frac{1}{2} \mathrm{i} (\sin \lambda  \cos \phi  (r_1+\mathrm{i} r_2 \cos \theta )+\sin \phi  (r_3 \sin \theta +\cos \lambda  (r_1 \cos \theta +\mathrm{i} r_2)))$ and $t^*$ is the conjugate of $t$. Therefore,
\begin{align}\label{eq21}
    &M_{dc}^t(\Theta) \notag \\
    &= \lVert  \bigtriangleup \Theta-\Theta \bigtriangleup\rVert_1 \notag \\
    &=\max_{\rho} \lVert  (\bigtriangleup \Theta-\Theta \bigtriangleup)\rho\rVert_1 \notag \\
    &=\max_{|\vec r|= 1} \frac{1}{2\sqrt{2}} \sqrt{(3+C_1 -C_3+4C_5)r_1^2+(5-C_1-C_3+4C_5)r_2^2+(2-2C_1)r_3^2+8(C_2+C_3)r_1 r_3} \notag \\
    &=\frac{1}{2\sqrt{2}} \sqrt{5-C_1-C_3+4C_5},
\end{align} where $C_1=2 \sin ^2\theta  \cos 2 \phi +\cos 2 \theta $, $C_2=\sin 2 \theta  \cos \lambda  \sin ^2\phi $, $C_3=\cos 2 \lambda (-2 \cos 2 \theta  \sin ^2\phi  +3 \cos 2 \phi +1)$, $C_4=\sin \theta  \sin \lambda  \sin 2 \phi$ and $C_5=\cos \theta  \sin 2 \lambda  \sin 2 \phi $.
It is found that the maximum value of Eq. (\ref{eq21}) is 1, that
is, $M_{dc}^t(\Theta)\le 1$ for any qubit unitary gate $\Theta$. By
fixing $\lambda=\frac{\pi}{2}$ in Eq. (\ref{eq21}), we obtain
\begin{align} \label{eq22}
    M_{dc}^t(\hat{\Theta})=\frac{1}{2} \sqrt{3-2 \cos 2 \theta  \sin ^2 \phi +\cos 2
    \phi}.
\end{align} And the surface of $M_{dc}^t(\hat{\Theta})$ in Eq. (\ref{eq22}) in which $\hat{\Theta}$ is induced by $U(\theta,\phi,\frac{\pi}{2})$ is plotted in Figure \ref{fig:Fig2}.
In particular, the qubit unitary gate $U(\theta,\phi,\frac{\pi}{2})$
degrade to certain quantum gates for specific $\theta$ and $\phi$,
and the corresponding quantity in Eq. (\ref{eq22}) can be
calculated, which are shown in Table 1.

\begin{table}[H]
    \centering
    \caption{The values of $M_{dc}^t(\hat{\Theta})$ for different operations induced by typical quantum gates}
    \resizebox{1.0\linewidth}{!}{
        \begin{tabular}{@{}c *{6}{>{\centering\arraybackslash}p{1.8cm}}@{}}
            \toprule
            gate & $\sigma_y$ & $\sigma_z$ & $S$ & $T$ & $R_x(\alpha)$ & $R_z(\alpha-\frac{\pi}{2})$ \\
            \midrule
            $\theta$ & $\pi$ & 0 & 0 & 0 & $\alpha$ & 0 \\
            \cline{1-7}
            $\phi$ & $\frac{\pi}{2}$ & $\frac{\pi}{2}$ & 0 & $-\frac{\pi}{4}$ & $-\frac{\pi}{2}$ & $\alpha-\frac{\pi}{2}$ \\
            \cline{1-7}
            $M_{dc}^t(\hat{\Theta})$ & 0 & 0 & 1 & $\frac{\sqrt{2}}{2}$ & $|\sin \alpha|$ & $|\sin \alpha|$ \\
            \bottomrule
        \end{tabular}
    }
\end{table}

\noindent {\bf 6. Applications in quantum
information processing tasks}\\\hspace*{\fill}

In this section, we apply the imaginarity measures
of quantum operations to discuss some problems in quantum
information processing tasks. First of all, we consider quantum
channel discrimination tasks\cite{WATJ,TRRB,PMWJ}. Denote by
$\mathcal{C_{AB}}$ and $\mathcal{FC_{AB}}$ the sets of channels and
free channels from $\mathcal{L(H_A)}$ to $\mathcal{L(H_B)}$,
respectively, $\mathcal{F_A}$ and $\mathcal{F_B}$ the sets of real
states on $\mathcal{H_A}$ and $\mathcal{H_B}$, respectively, and
$\mathcal{FP_B}$ the set of free POVMs $E=\{E_1,E_2\}$ on
$\mathcal{H_B}$. Given two channels $\Theta$ and $\Phi$ and a shared
probe state $\rho$, the success probability of distinguishing
$\Theta$ and $\Phi$ by the probe state $\rho$ coincides with the
success probability of distinguishing the states $\Theta(\rho)$ and
$\Phi(\rho)$ as\cite{LLBK}
\begin{align}\label{eq23}
    p_{\mathrm{succ}}(\Theta,\Phi,\rho)=\max_{\{\Pi,\mathbf{I}-\Pi\}} \left\{  \frac{1}{2}\mathrm{tr}\left[\Pi\Theta(\rho)\right ]+\frac{1}{2}\mathrm{tr}\left[(\mathbf{I}-\Pi)\Phi(\rho)\right] \right\},
\end{align} where the maximum is taken over all POVMs $\{\Pi,\mathbf{I}-\Pi\}$. In addition, from the Holevo-Helstrom theorem\cite{Helstrom1976}, it holds that $P_{\mathrm{succ}}(\Theta,\Phi,\rho)=\frac{1}{2}+\frac{1}{2}\lVert\Theta(\rho)-\Phi(\rho)\rVert_1$.

Following Refs. \cite{LLBK,MPWJ,TRRB19}, we analogously define the following success probabilities. First, the success probability of distinguishing $\Theta$ and $\Phi$ by the sets of probe states $\mathcal{F_A}$ and $\mathcal{D(H_A)}$ are  defined by
\begin{align}\label{eq24}
        p_{\mathrm{succ}}(\Theta,\Phi,\mathcal{F_A})=\min_{\rho\in\mathcal{F_A}}p_{\mathrm{succ}}(\Theta,\Phi,\rho)
\end{align} and
\begin{align}\label{eq25}
    p_{\mathrm{succ}}(\Theta,\Phi,\mathcal{D(H_A)})=\min_{\rho\in\mathcal{D(H_A)}}p_{\mathrm{succ}}(\Theta,\Phi,\rho),
\end{align} respectively. On the other hand, the success probability of distinguishing $\Theta$ from the set of free channels $\mathcal{FC_{AB}}$ by the probe state $\rho$ is defined by
\begin{align}\label{eq26}
    p_{\mathrm{succ}}(\Theta,\mathcal{FC_{AB}},\rho)=\min_{\Phi\in\mathcal{FC_{AB}}}
    p_{\mathrm{succ}}(\Theta,\Phi,\rho),
\end{align} and the maximum success probability of distinguishing $\Theta$ from the set of free channels $\mathcal{FC_{AB}}$ by the set of free states $\mathcal{F_A}$ can be further defined
as
\begin{align}\label{eq27}
    p_{\mathrm{succ}}(\Theta,\mathcal{FC_{AB}},\mathcal{F_A})=\max_{\rho\in\mathcal{F_{A}}}p_{\mathrm{succ}}(\Theta,\mathcal{FC_{AB}},\rho).
\end{align}
\indent In particular, we denote by
$\tilde{p}_{\mathrm{succ}}(\cdot,\cdot,\cdot)$ the success
probabilities corresponding to Eqs. (\ref{eq23}-\ref{eq27}),
respectively, by only performing free POVMs
$\{\Pi,\mathbf{I}-\Pi\}\in \mathcal{FP_B}$. In general, we have
$\tilde{p}_{\mathrm{succ}}(\Theta,\Phi,\rho)\le
p_{\mathrm{succ}}(\Theta,\Phi,\rho)$ and
$p_{\mathrm{succ}}(\Theta,\Phi,\mathcal{D(H_A)}) \le
p_{\mathrm{succ}}(\Theta,\Phi,\mathcal{F_A})$. Now we derive
relations between the aforementioned success probabilities.

\indent {\bf Theorem 12} Let
$\Theta\in\mathcal{FC_{AB}}$, $\Phi\in\mathcal{FC_{AB}}$, and
$\rho\in\mathcal{F_A}$ be a probe state. Then we have
\begin{align}
    p_{\mathrm{succ}}(\Theta,\Phi,\rho)=\tilde{p}_{\mathrm{succ}}(\Theta,\Phi,\rho). \notag
\end{align}
\indent \textit {Proof}. Let
$\rho=\sum\limits_{i,j}\rho_{ij}|a_i\rangle \langle a_j|$,
$\Theta\left(|a_i\rangle \langle a_j|
\right)=\sum\limits_{k,l}\Theta_{k,l}^{i,j}|b_k\rangle \langle
b_l|$, $\Phi\left(|a_i\rangle \langle a_j|
\right)=\sum\limits_{k,l}\Phi_{k,l}^{i,j}|b_k\rangle \langle b_l|$
and $\Pi=\sum\limits_{s,t}\Pi_{st}|b_s\rangle \langle b_t|$. Then we
have
\begin{align}\label{eq28}
    \mathrm{tr}\left[\Pi\Theta(\rho)\right]&=\mathrm{tr}\left[ \sum_{i,j,k,l,s,t} \Pi_{st}\Theta_{k,l}^{i,j} \rho_{ij} |b_s\rangle \langle b_t |b_k \rangle \langle b_l|\right] \notag \\
    &=\mathrm{tr}\left[ \sum_{i,j,k,l,s} \Pi_{sk}\Theta_{k,l}^{i,j} \rho_{ij} |b_s \rangle \langle b_l|\right] \notag \\
    &= \sum_{i,j,k,l} \Pi_{lk} \Theta_{k,l}^{i,j} \rho_{ij}.
\end{align} Note that $\rho_{ij}$ and $\Theta_{k,l}^{i,j}\in\mathbb{R}$ for any $i,j,k$, and $l$ since $\rho\in\mathcal{F_A}$ and $\Theta\in\mathcal{FC_{AB}}$. Therefore, we get
\begin{align}
    \sum_{i,j,k,l} \Pi_{lk} \Theta_{k,l}^{i,j} \rho_{ij}=\sum_{i,j,k,l} \mathrm{Re}\left( \Pi_{lk}\right)  \Theta_{k,l}^{i,j} \rho_{ij}. \notag
\end{align} Similar arguments show that
\begin{align}
    \sum_{i,j,k,l} \Pi_{lk} \Phi_{k,l}^{i,j} \rho_{ij}=\sum_{i,j,k,l} \mathrm{Re}\left( \Pi_{lk}\right)  \Phi_{k,l}^{i,j} \rho_{ij}. \notag
\end{align}

This means that for any POVM $\{\Pi,\mathbf{I}-\Pi\}$, we have
\begin{align}
 \frac{1}{2}\mathrm{tr}\left[\Pi\Theta(\rho)\right ]+\frac{1}{2}\mathrm{tr}\left[(\mathbf{I}-\Pi)\Phi(\rho)\right]=\frac{1}{2}\mathrm{tr}\left[\tilde{\Pi}\Theta(\rho)\right ]+\frac{1}{2}\mathrm{tr}\left[\left(\mathbf{I}-\tilde{\Pi}\right)\Phi(\rho)\right],\notag
\end{align} where $\tilde{\Pi}=\mathrm{Re}\left( \Pi\right)$.
Combining this with Eq. (\ref{eq23}), the conclusion follows.
$\hfill\qedsymbol$

The result indicates that if two channels to be
discriminated are priori known to be free and a real state $\rho$ is
used as the probe state, then the success probability of
distinguishing them can be obtained by performing only free POVMs,
which greatly reduces the cost required to determine the success
probability of discrimination and may be of significance in
practical experiments.

By Eq. (\ref{eq24}), it follows from Theorem 12 that
\begin{align}\label{eq29}
    p_{\mathrm{succ}}(\Theta,\Phi,\mathcal{F_A})=\tilde{p}_{\mathrm{succ}}(\Theta,\Phi,\mathcal{F_A}).
\end{align}
\indent {\bf Theorem 13} Let
$\Theta\in\mathcal{FC_{AB}}$ and $\Phi\in\mathcal{FC_{AB}}$. Then we
have
\begin{align}\label{eq30}
        \tilde{p}_{\mathrm{succ}}(\Theta,\Phi,\mathcal{F_A})=\tilde{p}_{\mathrm{succ}}(\Theta,\Phi,\mathcal{D(H_A)}).
\end{align}
\indent \textit {Proof}. It follows from the definition of
$\tilde{p}_{\mathrm{succ}}(\Theta,\Phi,\mathcal{D(H_A)})$ that
\begin{align}
    \tilde{p}_{\mathrm{succ}}(\Theta,\Phi,\mathcal{D(H_A)})=\min_{\rho\in\mathcal{D(H_A)}}\max_{\{\Pi,\mathbf{I}-\Pi\}\in\mathcal{FP_{B}}} \left\{  \frac{1}{2}\mathrm{tr}\left[\Pi\Theta(\rho)\right ]+\frac{1}{2}\mathrm{tr}\left[(\mathbf{I}-\Pi)\Phi(\rho)\right] \right\}. \notag
\end{align} From Eq. (\ref{eq28}), we have
\begin{align}
    \mathrm{tr}\left[\Pi\Theta(\rho)\right]=\sum_{i,j,k,l} \Pi_{lk} \Theta_{k,l}^{i,j} \rho_{ij}.\notag
\end{align} Since $\{\Pi,\mathbf{I}-\Pi\}\in\mathcal{FP_{B}}$ and $\Theta\in\mathcal{FC_{AB}}$, we have $\Pi_{lk},\Theta_{k,l}^{i,j}\in\mathbb{R}$ for all $i,j,k,l$, which yields that
\begin{align}
    \sum_{i,j,k,l} \Pi_{lk} \Theta_{k,l}^{i,j} \rho_{ij}=\sum_{i,j,k,l} \Pi_{lk} \Theta_{k,l}^{i,j} \mathrm{Re}\left(\rho_{ij} \right).\notag
\end{align} Similarly, one finds that
\begin{align}
\sum_{i,j,k,l} \Pi_{lk} \Phi_{k,l}^{i,j} \rho_{ij}=\sum_{i,j,k,l} \Pi_{lk} \Phi_{k,l}^{i,j} \mathrm{Re}\left(\rho_{ij} \right).\notag
\end{align} Thus we have
\begin{align}
     \frac{1}{2}\mathrm{tr}\left[\Pi\Theta(\rho)\right ]+\frac{1}{2}\mathrm{tr}\left[\left(\mathbf{I}-\Pi\right)\Phi(\rho)\right]=\frac{1}{2}\mathrm{tr}\left[\tilde{\Pi}\Theta(\bigtriangleup(\rho))\right ]+\frac{1}{2}\mathrm{tr}\left[\left(\mathbf{I}-\tilde{\Pi}\right)\Phi(\bigtriangleup(\rho))\right],\notag
\end{align}
where $\tilde{\Pi}=\mathrm{Re}\left( \Pi\right)$, and so
\begin{align}
    &\min_{\rho\in\mathcal{D(H_A)}}\max_{\{\Pi,\mathbf{I}-\Pi\}\in\mathcal{FP_{B}}} \left\{  \frac{1}{2}\mathrm{tr}\left[\Pi\Theta(\rho)\right ]+\frac{1}{2}\mathrm{tr}\left[\left(\mathbf{I}-\Pi\right)\Phi(\rho)\right] \right\} \notag \\
    &=\min_{\rho\in\mathcal{F_A}}\max_{\{\Pi,\mathbf{I}-\Pi\}\in\mathcal{FP_{B}}} \left\{  \frac{1}{2}\mathrm{tr}\left[\Pi\Theta(\rho)\right ]+\frac{1}{2}\mathrm{tr}\left[\left(\mathbf{I}-\Pi\right)\Phi(\rho)\right] \right\}, \notag
\end{align} which implies that Eq. (\ref{eq30}) holds. $\hfill\qedsymbol$

It follows from Theorem 13 that when two candidate
channels are priori known to be free ones and only free POVMs are
performed, the success probability of distinguishing them by the set
of probe states in $\mathcal{D(H_A)}$ can be achieved using solely
the set of free states $\mathcal{F_A}$ as probe states. This means
that to get the success probability in this scenario, we do not need
any extra imaginarity resource in the probe states.

Combining Eqs. (\ref{eq29}) and (\ref{eq30}), it
follows that
\begin{align}\label{eq31}
    p_{\mathrm{succ}}(\Theta,\Phi,\mathcal{F_A})=\tilde{p}_{\mathrm{succ}}(\Theta,\Phi,\mathcal{D(H_A)}).
\end{align}
\indent This equality demonstrates that in quantum
channel discrimination for resourceless channels, if you want to
estimate the success probability in experiments by just using POVMs
without the imaginary resource, a price has to be paied for
utilizing all quantum states as probe states instead of the ones
without imaginarity resource, or if you want to evaluate the success
probability by using probe states as the ones with no imaginarity,
the measurements should go over all POVMs instead of the ones with
no imaginarity.

Now we present the
relations between $\tilde{M}_{c}^t(\Theta)$ and
$p_{\mathrm{succ}}(\Theta,\mathcal{FC_{AB}},\mathcal{F_A})$.

\indent {\bf Theorem 14} Let
$\Theta\in\mathcal{C_{AB}}$. Then we have
\begin{align}\label{eq32}
     p_{\mathrm{succ}}(\Theta,\mathcal{FC_{AB}},\mathcal{F_A}) \le \frac{1}{2}+\frac{1}{2} \tilde{M}_{c}^t(\Theta),
\end{align} where $\tilde{M}_{c}^t(\Theta)=\min\limits_{\Phi\in\mathcal{FC_{AB}}}\lVert\Theta\bigtriangleup-\Phi\bigtriangleup\rVert_1$.\\
\indent \textit {Proof}.
 First of all, note that
 \begin{align}\label{eq33}
    \min_{\Phi\in\mathcal{FC_{AB}}}\lVert\Theta(\rho)-\Phi(\rho)\rVert_1\le
    \lVert\Theta(\rho)-\Phi(\rho)\rVert_1,
 \end{align}
where $\Phi$ on the right hand side of Eq. (\ref{eq33}) is any
element in $\mathcal{FC_{AB}}$. Taking the maximum over
$\rho\in\mathcal{D(H_A)}$ on both sides of Eq. (\ref{eq33}), we
obtain
 \begin{align}\label{eq34}
    \max_{\rho\in\mathcal{D(H_A)}}\min_{\Phi\in\mathcal{FC_{AB}}}\lVert\Theta(\rho)-\Phi(\rho)\rVert_1\le
    \max_{\rho\in\mathcal{D(H_A)}}\lVert\Theta(\rho)-\Phi(\rho)\rVert_1,
 \end{align} which implies that
 \begin{align}\label{eq35}
    \max_{\rho\in\mathcal{D(H_A)}}\min_{\Phi\in\mathcal{FC_{AB}}}\lVert\Theta(\rho)-\Phi(\rho)\rVert_1\le \min_{\Phi\in\mathcal{FC_{AB}}}\max_{\rho\in\mathcal{D(H_A)}}\lVert\Theta(\rho)-\Phi(\rho)\rVert_1.
 \end{align} Therefore, we have
 \begin{align}
    &p_{\mathrm{succ}}(\Theta,\mathcal{FC_{AB}},\mathcal{F_A})\notag \\
    &=\max_{\rho\in\mathcal{F_A}}\min_{\Phi\in \mathcal{FC_{AB}}}\left\{ \frac{1}{2}+\frac{1}{2}\lVert\Theta(\rho)-\Phi(\rho)\rVert_1 \right\}\notag\\
    &\le\min_{\Phi\in \mathcal{FC_{AB}}}\max_{\rho\in\mathcal{F_A}} \left\{ \frac{1}{2}+\frac{1}{2}\lVert\Theta(\rho)-\Phi(\rho)\rVert_1 \right\}\notag\\
    &=\min_{\Phi\in \mathcal{FC_{AB}}}\max_{\rho\in\mathcal{D(H_A)}} \left\{ \frac{1}{2}+\frac{1}{2}\lVert\Theta\bigtriangleup(\rho)-\Phi\bigtriangleup(\rho)\rVert_1\right\} \notag\\
    &=\frac{1}{2}+\frac{1}{2} \tilde{M}_{c}^t(\Theta),\notag
 \end{align}
where the inequality follows from Eq. (\ref{eq35}).
$\hfill\qedsymbol$

 Theorem 14 shows that the maximum success probability $p_{\mathrm{succ}}(\Theta,\mathcal{FC_{AB}},\mathcal{F_A})$ is upper bounded by $\frac{1}{2}+\frac{1}{2} \tilde{M}_{c}^t(\Theta)$.
Combining Eq. (\ref{eq32}) with item (1) in Theorem 10, it follows
that $$p_{\mathrm{succ}}(\Theta,\mathcal{FC_{AB}},\mathcal{F_A})\le
\frac{1}{2}+\frac{1}{2}\tilde{\mathscr{I}}_w(\Theta),$$ where
$\tilde{\mathscr{I}}_w(\Theta)=\min\left \{ 0\le s\le1: \Theta=(1-s)
\Phi +s\Lambda, \Phi \in \mathcal{FC_{AB}},\Lambda \in
\mathcal{C_{AB}} \right \}$, indicating that the weight of
imaginarity can also be employed to give an upper bound of the
success probability.

Let $\Theta\in\mathcal{C_{AB}}$ and
$\Phi\in\mathcal{FC_{BC}}$. Since $\tilde{M}_{c}^t(\Theta)$
satisfies (M2a), the maximum success probability of distinguishing
$\Phi\circ\Theta$ from the set of free channels $\mathcal{FC_{AC}}$
by the set of free states $\mathcal{F_A}$ satisfies
 \begin{align} p_{\mathrm{succ}}(\Phi\circ\Theta,\mathcal{FC_{AC}},\mathcal{F_A}) \le \frac{1}{2}+\frac{1}{2} \tilde{M}_{c}^t(\Phi\circ\Theta)\le \frac{1}{2}+\frac{1}{2}\tilde{M}_{c}^t(\Theta). \notag
 \end{align}
Since $\tilde{M}_{c}^t(\Theta)$ also satisfies (M2b), similar result
holds for right composition. Also,
$\frac{1}{2}+\frac{1}{2}\tilde{\mathscr{I}}_w(\Theta)$ serves as the
upper bounds. This demonstrates if the channel is
preprocessed/postprocessed with a free channel, the maximum success
probability can be upper bounded by the same quantity.

Let $\Theta\in\mathcal{C_{AB}}$ and
$\Phi\in\mathcal{FC_{CD}}$. Then with the help of item (3) in
Theorem 10, the maximum success probability
$p_{\mathrm{succ}}(\Phi\otimes\Theta,\mathcal{FC_{CADB}},\mathcal{F_{CA}})$
satisfies
 \begin{align}
    p_{\mathrm{succ}}(\Phi\otimes\Theta,\mathcal{FC_{CADB}},\mathcal{F_{CA}})\le\frac{1}{2}+\frac{1}{2} \tilde{M}_{c}^t(\Phi\otimes\Theta)\le\frac{1}{2}+\frac{1}{2}\tilde{\mathscr{I}}_w(\Theta).\notag
 \end{align}
Since $\tilde{M}_{c}^t(\Theta)$ does not satisfy monotonicity under
tensor product in general, we can only assert that it is upper
bounded by $\frac{1}{2}+\frac{1}{2}\tilde{\mathscr{I}}_w(\Theta)$
instead of $\frac{1}{2}+\frac{1}{2} \tilde{M}_{c}^t(\Theta)$. The
above argument still holds for
$p_{\mathrm{succ}}(\Theta\otimes\Phi,\mathcal{FC_{ACBD}},\mathcal{F_{AC}})$
for $\Theta\in\mathcal{C_{AB}}$ and $\Phi\in\mathcal{FC_{CD}}$.

Next consider a class of entanglement-assisted state
exclusion tasks\cite{URBT,DASP,YMLY}. In this scenario, we first
impose the quantum channel $\mathbb{I}\otimes\Theta$ on the state
ensemble $\{p_i,\rho_i\}$, where $\rho_i\in\mathcal{D(H_{AB})}$, and
then implement the measurement $\{M_i\}$ on the output system. Then
the average error probability for this task is given by\cite{YMLY}
 \begin{align}\label{eq36}
    p_{\mathrm{err}}\left( \{p_i,\rho_i\},\{M_i\},\mathbb{I}\otimes\Theta\right)=\sum_i p_i \mathrm{tr}\left[M_i\left((\mathbb{I}\otimes\Theta)(\rho_i)\right) \right].
 \end{align} Since $\mathcal{FC_{AB}}$ is a convex and compact set, by Theorem 4 in Ref. \cite{YMLY}, we obtain that
 \begin{align}\label{eq37}
    \min_{\{p_i,\rho_i\},\{M_i\}}\frac{ p_{\mathrm{err}}\left( \{p_i,\rho_i\},\{M_i\},\mathbb{I}\otimes\Theta\right) }{ \min\limits_{\Phi\in\mathcal{FC_{AB}}}p_{\mathrm{err}}\left( \{p_i,\rho_i\},\{M_i\},\mathbb{I}\otimes\Phi\right)}=1-\tilde{\mathscr{I}}_w(\Theta),
 \end{align} where $ \tilde{\mathscr{I}}_w(\Theta)=\min\left \{ 0\le s\le1: \Theta=(1-s) \Phi +s\Lambda, {\rm{\ for \ some\ }} \Phi \in \mathcal{FC_{AB}} {\rm{\ and\ }} \Lambda \in \mathcal{C_{AB}} \right \}$.\\

\noindent {\bf 7. Conclusions}\\\hspace*{\fill}\\
\indent Inspired by \cite{TTED}, in this work, we defined three
kinds of free operations in imaginarity resource theory through the
ability to create or detect imaginarity, i.e., detection real
operations, creation real operations, and creation detection real
operations, and proved that they are equivalent. We built the
connection between these operations and their associated Kraus
operators, and clarified the relations among our concepts of free
operations and existing ones. Based on this, we further introduced
free superoperations in imaginarity resource theory, and established
a new framework for imaginarity of quantum operations. Under this
framework, we have proposed some imaginarity measures of quantum
operations $M_c(\Theta)$, $M_d(\Theta)$, $M_{dc}(\Theta)$ based on
norms and the weight of imaginarity of quantum operations
$\mathscr{I}_w(\Theta)$, and investigated their properties.
Moreover, the relations among $M_c(\Theta)$, $M_d(\Theta)$ and
$M_{dc}(\Theta)$, as well as between $M_{dc}(\Theta)$ and $
\mathscr{I}_w(\Theta)$, are all explicitly given. Besides, we have
provided the applications of our results in two kinds of information
processing tasks, that is, channel discrimination and
entanglement-assisted exclusion, by establishing the relations
between the imaginarity measure of quantum operations and the
success probability of discrimination/the average error probability
of exclusion. Our results may shed some new light on the study of
quantifying imaginarity of quantum operations, and help to
understand the essence of imaginarity from the dynamical
perspective.

\vskip0.1in

\noindent

\subsubsection*{Acknowledgements}
This work was supported by National Natural Science Foundation of China (Grant Nos. 12561084, 12161056) and Natural Science Foundation of Jiangxi Province (Grant No. 20232ACB211003).\\
\subsubsection*{Conflict of interest} The authors declare that they have no conflict of
interest.

\small {}

\end{document}